\documentclass[dvips]{acta}
\usepackage{supertabular,lscape,epsfig}
\usepackage{amssymb}
\usepackage{amsmath}
\DeclareSymbolFont{ppa}{OT1}{ppl}{m}{it}
\DeclareMathSymbol{\vv}{\mathalpha}{ppa}{'166}

\newfont{\hb}{rphvb at 10pt}
\newfont{\hbo}{rphvbo at 10pt}
\newfont{\bitt}{rptmbi at 12pt}
\newfont{\bits}{rptmbi at 11pt}

\SetPages{247}{268}

\SetVol{62}{2012}

\begin{document}

\newcommand{\TabApp}[2]{\begin{center}\parbox[t]{#1}{\centerline{
  {\bf Appendix}}
  \vskip2mm
  \centerline{\small {\spaceskip 2pt plus 1pt minus 1pt T a b l e}
  \refstepcounter{table}\thetable}
  \vskip2mm
  \centerline{\footnotesize #2}}
  \vskip3mm
\end{center}}

\newcommand{\TabCapp}[2]{\begin{center}\parbox[t]{#1}{\centerline{
  \small {\spaceskip 2pt plus 1pt minus 1pt T a b l e}
  \refstepcounter{table}\thetable}
  \vskip2mm
  \centerline{\footnotesize #2}}
  \vskip3mm
\end{center}}

\newcommand{\TTabCap}[3]{\begin{center}\parbox[t]{#1}{\centerline{
  \small {\spaceskip 2pt plus 1pt minus 1pt T a b l e}
  \refstepcounter{table}\thetable}
  \vskip2mm
  \centerline{\footnotesize #2}
  \centerline{\footnotesize #3}}
  \vskip1mm
\end{center}}

\newcommand{\MakeTableApp}[4]{\begin{table}[p]\TabApp{#2}{#3}
  \begin{center} \TableFont \begin{tabular}{#1} #4 
  \end{tabular}\end{center}\end{table}}

\newcommand{\MakeTableSepp}[4]{\begin{table}[p]\TabCapp{#2}{#3}
  \begin{center} \TableFont \begin{tabular}{#1} #4 
  \end{tabular}\end{center}\end{table}}

\newcommand{\MakeTableee}[4]{\begin{table}[htb]\TabCapp{#2}{#3}
  \begin{center} \TableFont \begin{tabular}{#1} #4
  \end{tabular}\end{center}\end{table}}

\newcommand{\MakeTablee}[5]{\begin{table}[htb]\TTabCap{#2}{#3}{#4}
  \begin{center} \TableFont \begin{tabular}{#1} #5 
  \end{tabular}\end{center}\end{table}}

\newfont{\bb}{ptmbi8t at 12pt}
\newfont{\bbb}{cmbxti10}
\newfont{\bbbb}{cmbxti10 at 9pt}
\newcommand{\uprule}{\rule{0pt}{2.5ex}}
\newcommand{\douprule}{\rule[-2ex]{0pt}{4.5ex}}
\newcommand{\dorule}{\rule[-2ex]{0pt}{2ex}}
\def\thefootnote{\fnsymbol{footnote}}
\begin{Titlepage}
\Title{Photometric Maps Based on the OGLE-III Shallow Survey\\
in the Large Magellanic Cloud\footnote{Based on observations obtained with
the 1.3 m Warsaw telescope at the Las Campanas Observatory of the Carnegie
Institution for Science.}}
\vspace*{7pt}
\Author{K.~~U~l~a~c~z~y~k$^1$,~~ M.\,K.~~S~z~y~m~a~ñ~s~k~i$^1$,~~ A.~~U~d~a~l~s~k~i$^1$,~~ \\
M.~~K~u~b~i~a~k$^1$,~~ G.~~P~i~e~t~r~z~y~ñ~s~k~i$^{1,2}$,~~ I.~~S~o~s~z~y~ñ~s~k~i$^1$,\\
£.~~W~y~r~z~y~k~o~w~s~k~i$^{1,3}$,~~ R.~~P~o~l~e~s~k~i$^1$, \\
W.~~G~i~e~r~e~n$^2$,~~ A.\,R.~~W~a~l~k~e~r$^4$ ~~and~~ A.~~G~a~r~c~i~a~-~V~a~r~e~l~a$^5$}
{$^1$Warsaw University Observatory, Al.~Ujazdowskie~4, 00-478~Warszawa, Poland\\
e-mail: (kulaczyk,msz,udalski,mk,pietrzyn,soszynsk,wyrzykow,rpoleski)@astrouw.edu.pl\\
\vskip3pt
$^2$Universidad de Concepción, Departamento de Astronomia, Casilla 160-C, Concepción, Chile\\
e-mail: wgieren@astro-udec.cl\\
\vskip3pt
$^3$Institute of Astronomy, University of Cambridge, Madingley Road, Cambridge CB3~0HA, UK\\
\vskip3pt
$^4$Cerro Tololo Inter-American Observatory, Casilla~603, La~Serena, Chile\\
e-mail: awalker@ctio.noao.edu\\
\vskip3pt
$^5$Departamento de Fisica, Universidad de los Andes, Bogota, Colombia\\
e-mail: josegarc@uniandes.edu.co} 

\vspace*{7pt}
\Received{October 1, 2012}
\end{Titlepage}

\vspace*{3pt}
\Abstract{
We present photometric maps based on data from the shallow survey in the
Large Magellanic Cloud performed as the supplementary project during the
third phase of the Optical Gravitational Lensing Experiment. They cover
about 40 square degrees in the LMC and contain mean calibrated {\it VI}
photometry and astrometry of about 1.7 million stars. The magnitudes of
the registered objects range from 9.1 to 18.5.

We discuss the quality of data and present color--magnitude diagrams of
selected fields. The maps together with the main LMC photometric maps are
available to the astronomical community from the OGLE Internet archive.
}{Magellanic Clouds -- Surveys -- Catalogs -- Techniques: photometric}

\vspace*{7pt}
\Section{Introduction}
The extensive OGLE-III Photometric Maps of the Large Magellanic Cloud have
already been published in Udalski \etal (2008b). They contained stars as
faint as 20.5~mag in {\it I}-band and showed high level of completeness.
However, these maps suffered from obvious limitation -- very bright objects
became saturated during regular observations (performed with exposure times
equal to 180~s and 240~s in {\it I}- and {\it V}-band, respectively).
Typically, overexposure took place at $12.5\div13$~mag in the {\it I}-band,
depending on stellar density. Obviously, numerous fainter stars registered
on images in the vicinity of saturated objects could also be affected.

The main purpose of the shallow survey described in this paper was to
complement the OGLE-III databases for the studied fields with objects for
which photometry had not been previously obtained due to saturation in
single or both bands. Here we describe reduction procedures applied to the
collected short-exposure images, we present photometric maps for the Large
Magellanic Cloud and we discuss quality of the data.

\renewcommand{\arraystretch}{0.97}
\MakeTableee{|c|c|c||c|c|c||c|c|c||c|c|c|}{12.5cm}{Number of observations per field}
{
\hline
\douprule Field & $N_I$ & $N_V$ & Field & $N_I$ & $N_V$ & Field & $N_I$ & $N_V$ & Field & $N_I$ & $N_V$\\
\hline
LMC100&65&63&LMC131&22&22&LMC157&25&25&LMC190&20&20\\
LMC101&59&58&LMC132&21&21&LMC158&25&24&LMC191&51&51\\
LMC102&57&58&LMC133&86&83&LMC159&63&64&LMC192&51&50\\
LMC103&57&56&LMC134&81&80&LMC160&63&62&LMC193&50&50\\
LMC104&54&53&LMC135&80&78&LMC161&61&61&LMC194&20&20\\
LMC105&52&52&LMC136&78&78&LMC162&57&57&LMC195&20&20\\
LMC107&78&78&LMC137&21&21&LMC163&57&54&LMC196&21&21\\
LMC108&77&77&LMC138&26&26&LMC164&55&55&LMC197&21&21\\
LMC109&74&75&LMC139&26&26&LMC165&54&54&LMC198&21&21\\
LMC110&75&75&LMC140&26&25&LMC167&60&59&LMC199&21&21\\
LMC111&73&72&LMC141&25&25&LMC168&57&57&LMC200&21&20\\
LMC112&69&70&LMC142&25&25&LMC169&57&56&LMC201&19&19\\
LMC113&70&69&LMC143&25&25&LMC170&55&55&LMC202&19&19\\
LMC114&68&67&LMC144&25&25&LMC171&55&55&LMC203&19&19\\
LMC117&72&73&LMC145&26&26&LMC172&53&52&LMC204&17&17\\
LMC118&73&73&LMC146&26&26&LMC174&59&60&LMC205&17&17\\
LMC119&72&72&LMC147&26&26&LMC175&59&57&LMC206&17&17\\
LMC120&72&72&LMC148&25&25&LMC176&59&59&LMC207&15&15\\
LMC121&71&69&LMC149&25&24&LMC177&55&55&LMC208&15&15\\
LMC124&82&81&LMC150&25&25&LMC178&58&57&LMC209&14&14\\
LMC125&81&80&LMC151&25&25&LMC179&58&56&LMC210&14&14\\
LMC126&80&78&LMC152&30&30&LMC183&63&62&LMC211&14&14\\
LMC127&78&77&LMC153&28&28&LMC184&58&57&LMC212&13&13\\
LMC128&75&75&LMC154&26&25&LMC185&58&58&LMC213&13&13\\
LMC129&72&72&LMC155&26&26&LMC186&56&56&LMC214&13&13\\
LMC130&69&69&LMC156&26&26&LMC189&20&20&LMC215&13&12\\
\hline
}

\Section{Observations}
Observational data were collected using 1.3-m Warsaw Telescope located at
Las Campanas Observatory, operated by the Carnegie Institution for Science.
During the third phase of the OGLE project the telescope was equipped with
the eight chip mosaic camera covering approximately $35\arcm\times 35\arcm$
field of view with the scale of 0.26 arcsec/pixel. The {\it I} and {\it V}
filters used in the survey closely resembled standard filters, although
{\it I}-band filter had higher transmission for longer
wavelengths. Detailed description of the whole instrumentation can be found
in Udalski (2003). We also used data from OGLE-IV survey to correct the
photometry of red objects.

Images were registered during nights of seeing worse then about 2\arcs. It
allowed to blur profiles of the brightest stars preventing from the
saturation. Moreover we were able to perform observations during nights of
poor weather which normally would be lost for the main OGLE survey.

Typical exposure time was equal to 15~s for both filters. A process of
camera shutter opening took about 0.3~s so the possible shutter error
was negligible. For a period of 4.5~years total amount of 74\,192
subfield images were registered through both filters. Table~1 and diagram
in Fig.~1 present number of observations available for each field. The
standard OGLE-III field designations were used as described in
Udalski \etal (2008b).
\begin{figure}[h]
\centerline{\includegraphics[width=10.7cm, bb=0 170 570 700]{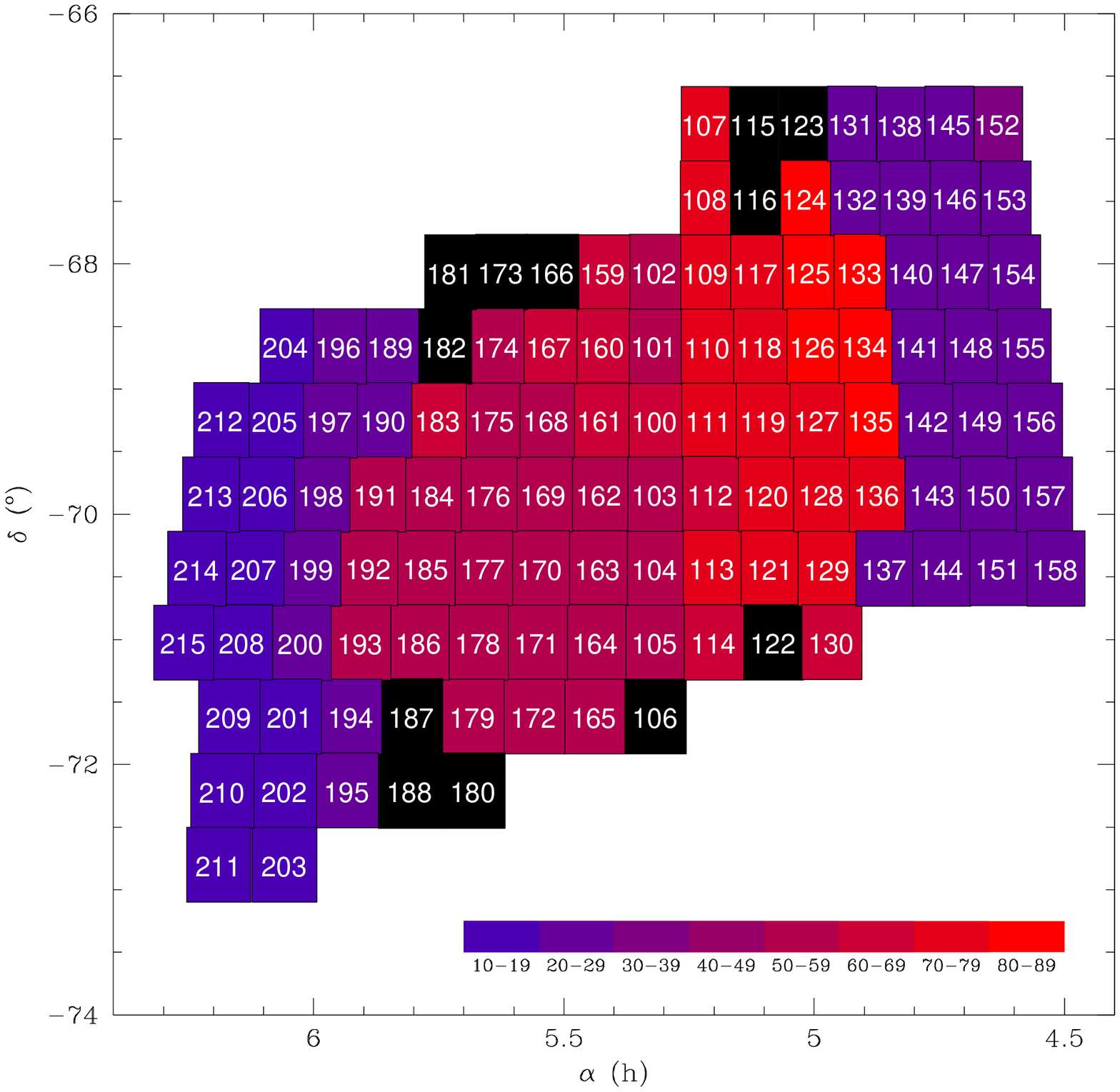}}
\FigCap{Number of observations per field. The black color 
means ``no observations''.}
\end{figure}

\Section{Data Reductions}
Preliminary reductions of all images -- debiasing and flatfielding -- were
performed immediately after collecting at the telescope using the standard
OGLE pipelines (Udalski 2003).

In order to minimize PSF variability each subfield image (corresponding to
one camera chip) was divided into two parts with overlapping region of 40
pixels.

Then profile photometry with {\sc DaoPhot/Allstar} package (Stetson 1987)
was performed for all obtained $2048\times2068$ images. PSF profiles were
derived from at least 15 isolated stars selected from initial list of 30
objects filtered by rejecting 3$\sigma$ deviations from the model. For all
images database was constructed consisting of photometric measurements in
the {\it I}- and {\it V}-band as well as a position data.

The linear shifts in $(X,Y)$ coordinates between subsequent images of the
same photometric subfield were derived and for all detected objects light
curves were prepared. The small inaccuracies of the telescope pointing
allowed to cover gaps between neighboring chips and fields. Median of the
absolute values of the shifts were equal to 28.5 pixels in $X$ axis and
29.2 pixels in $Y$ axis for {\it I} filter (relative to selected image of
given field with the best seeing). Corresponding values for {\it V} filter
were equal to 31.5 and 25.4. Maximum shift values were equal to (220, 340)
pixels in the {\it I}-band and (250, 240) pixels in the {\it V}-band.

All photometric measurements were corrected for chip ``flatness'' based on
observations of standard Landolt (1992) stars set at various chip
coordinates according to the description in Udalski \etal (2008a). Average
number of observations per object is equal to 45 in each band but it varies
significantly from field to field as shown in Fig.~1.

\begin{figure}[htb]
\centerline{\includegraphics[width=7cm]{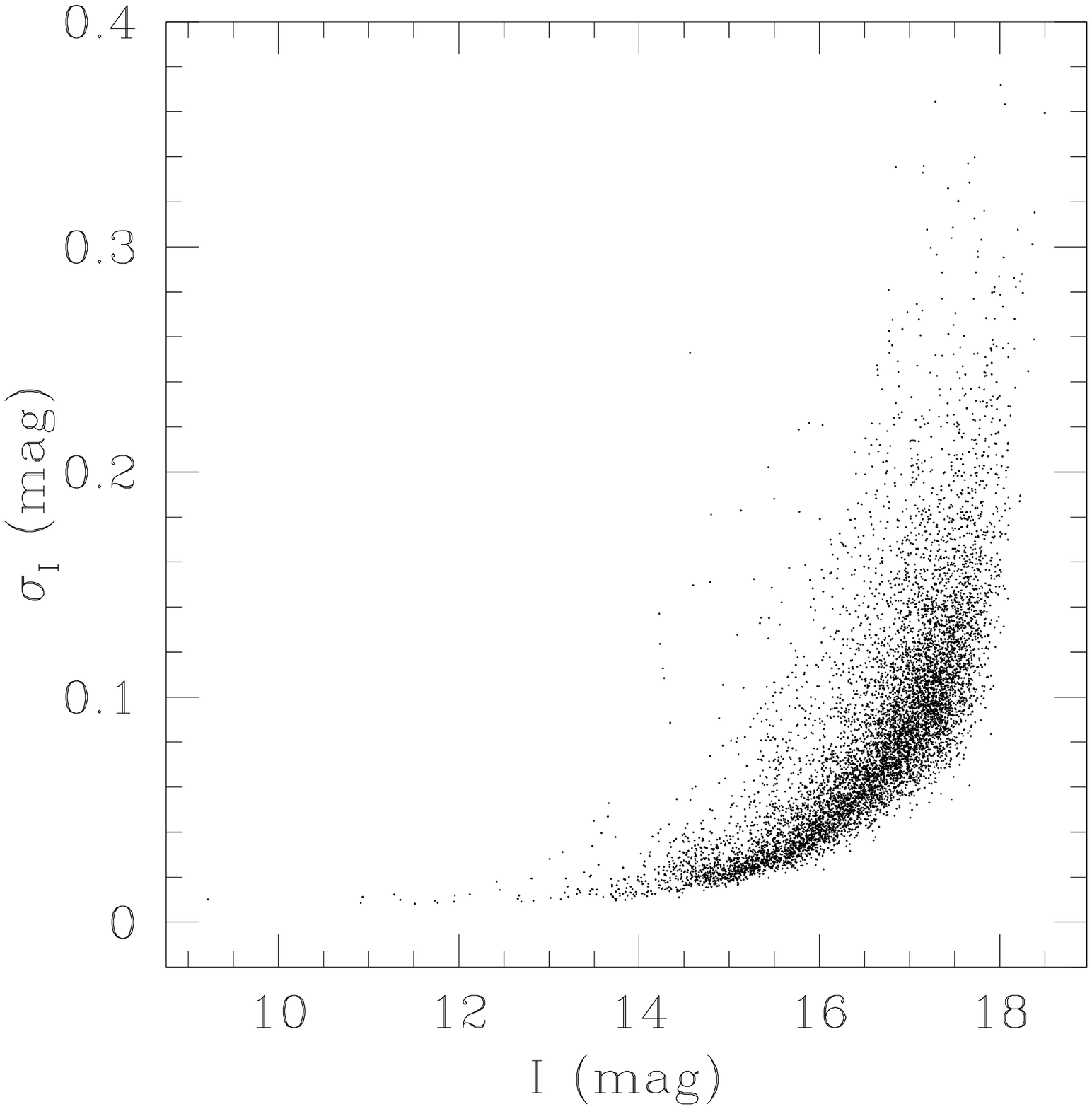}\hfill\includegraphics[width=7cm]{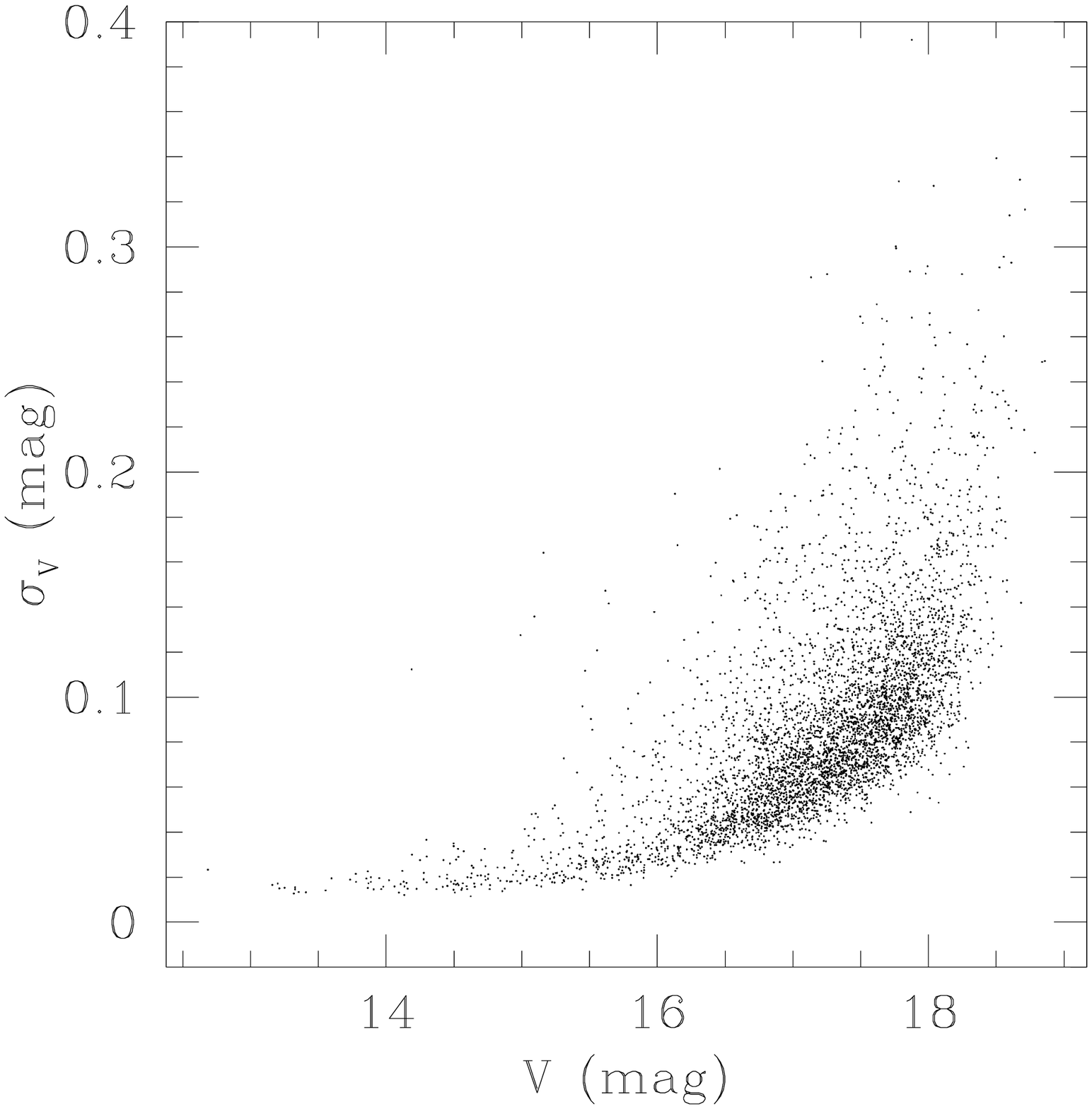}}
\vskip5pt
\FigCap{Standard deviation of magnitudes for the LMC100.1 subfield}
\end{figure}
Photometric databases were constructed separately for each subfield
(corresponding to one chip image) so some stars are present in multiple
subfields.

\begin{figure}[htb]
\centerline{\includegraphics[width=7cm]{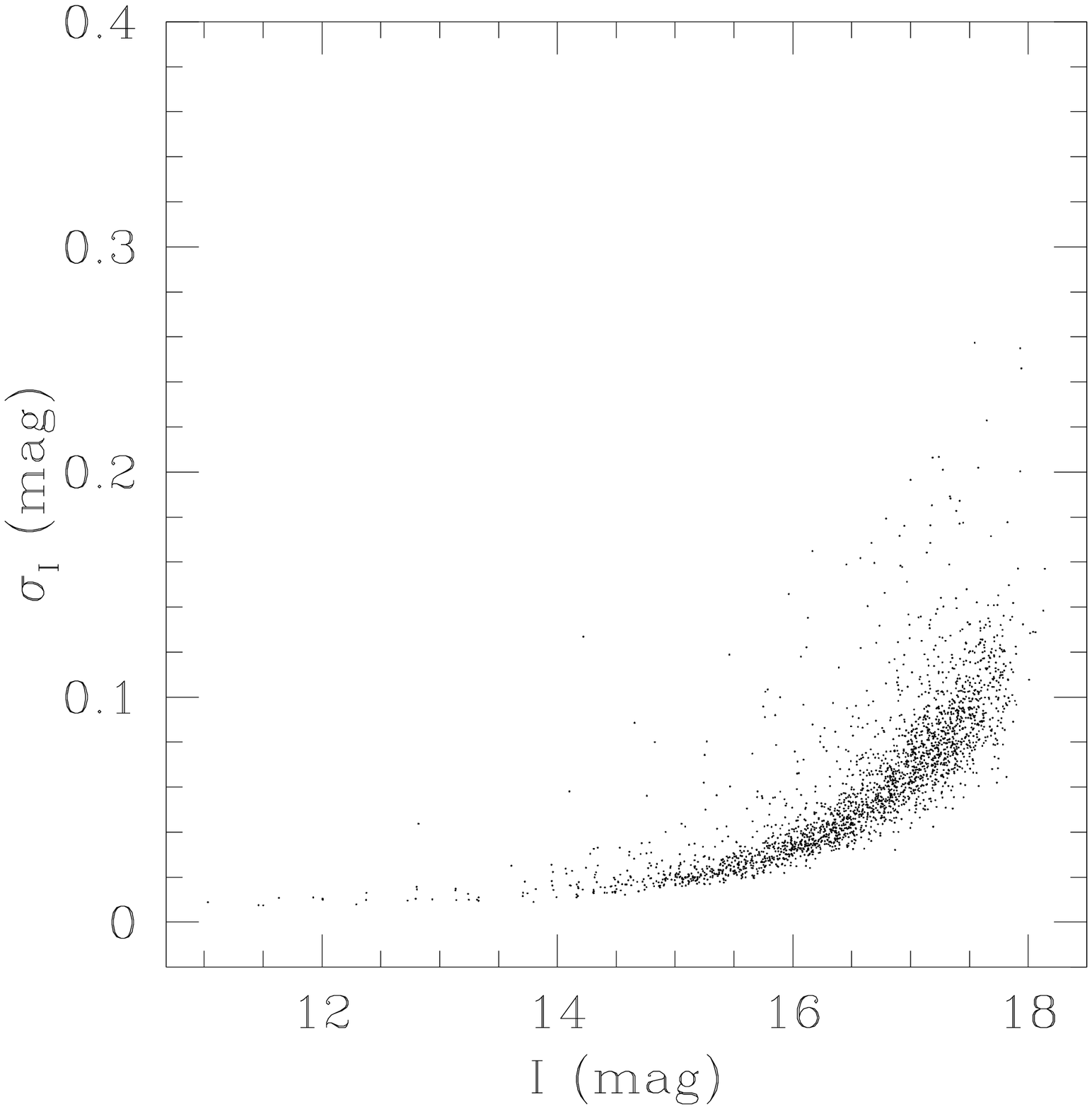}\hfill\includegraphics[width=7cm]{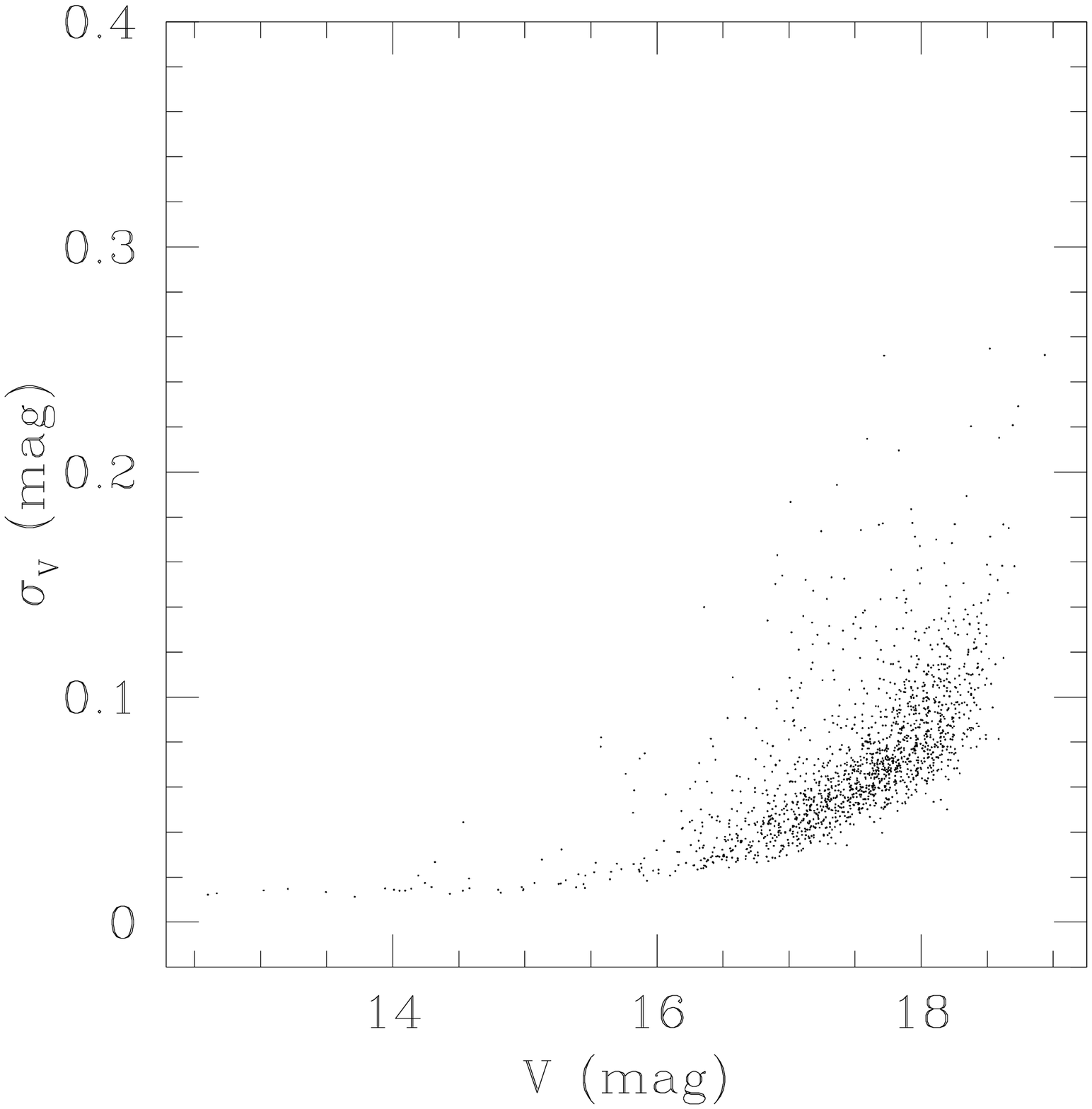}}
\vskip5pt
\FigCap{Standard deviation of magnitudes for the LMC185.2 subfield}
\end{figure}
In order to check the quality of photometry we constructed diagrams shown
in Figs.~2 and 3 presenting deviation of magnitudes as a function of
magnitude for observations through both filters. The main goal of the
survey was measuring stars brighter than 13~mag and as one can see the mean
dispersion in {\it I}-band for such objects is significantly smaller than
0.02~mag.

\Section{Photometric and Astrometric Calibration}
The same fields were observed as in the main OGLE-III survey so we were
able to use the calibrated photometry and astrometry as a reference.

For each photometric subfield ($2048\times2068$ pixels) the median value of
photometric shifts for constant stars cross-correlated with the regular
OGLE-III data was calculated with iterative 3$\sigma$ clipping and all
photometric measurements were appropriately adjusted. It was performed
independently for observations in {\it I}- and {\it V}-bands. The accuracy
of a zero-level point is within range 0.02--0.04 mag and shows dependency on
star density in a given field. Figs.~4 and 5 present difference between
standard OGLE-III photometry and shallow survey photometry as a function of
chip coordinates -- no significant correlation is visible.
\begin{figure}[p]
\includegraphics[width=9.5cm]{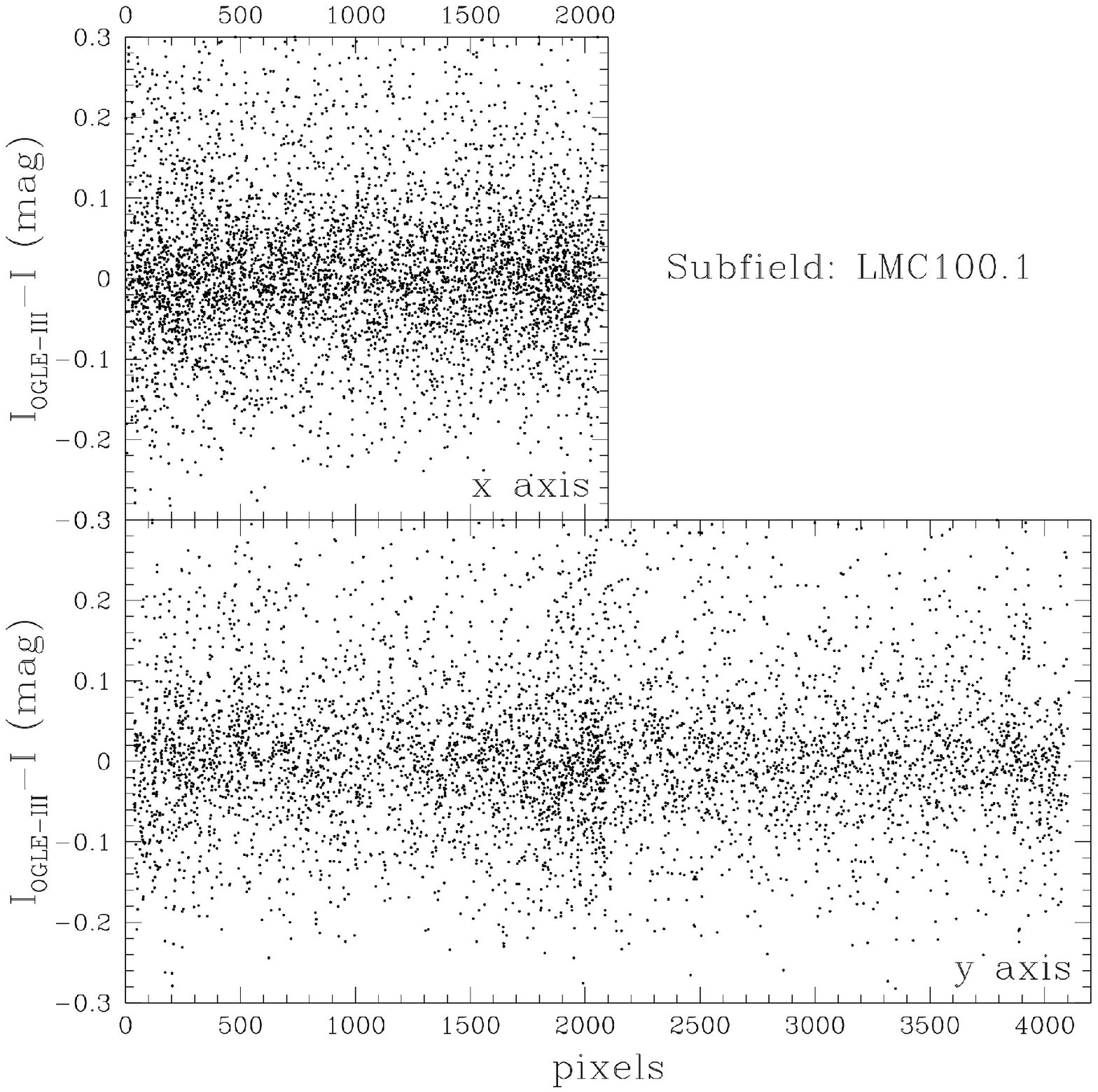}
\includegraphics[width=9.5cm]{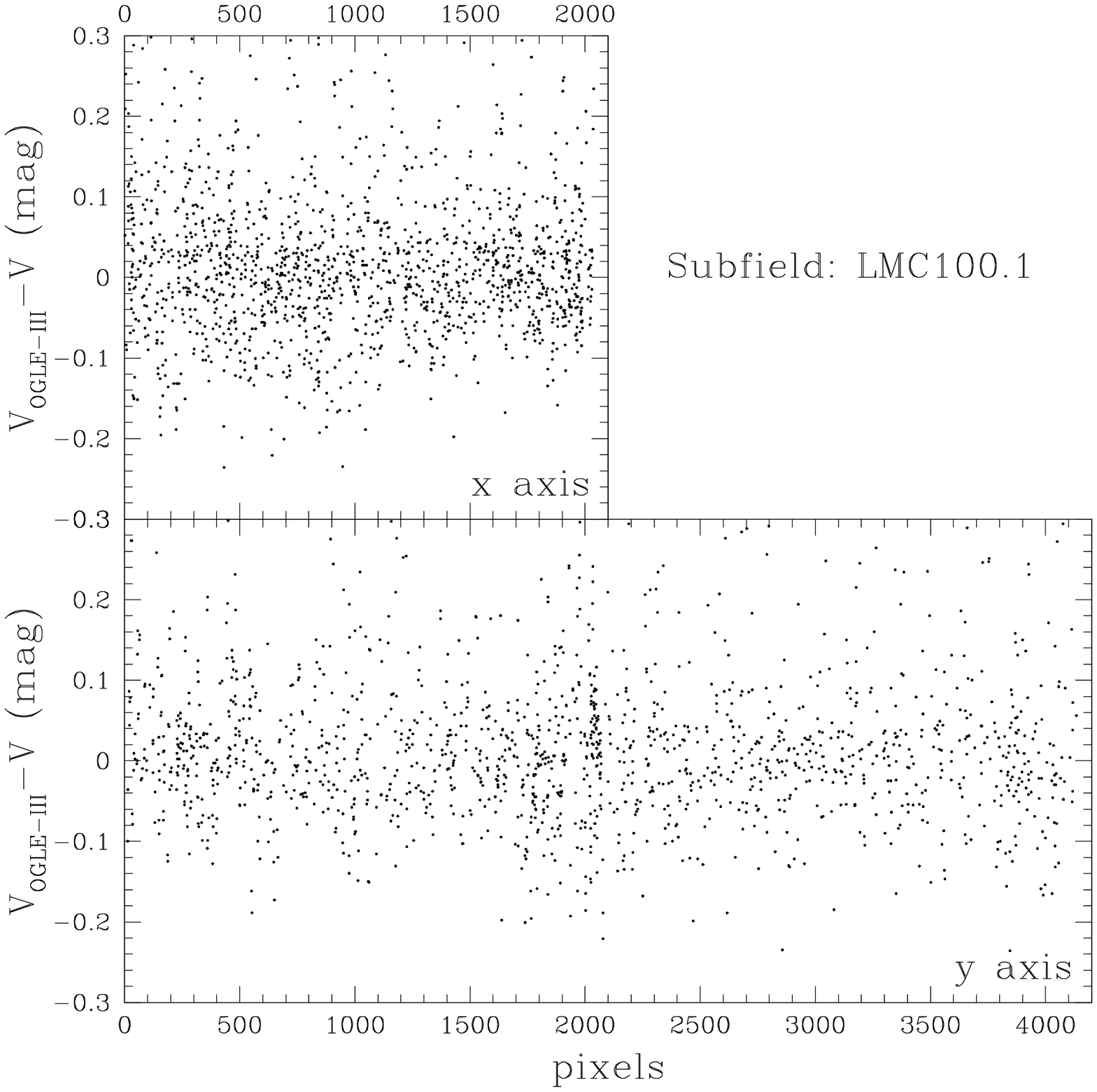}
\FigCap{Difference between standard OGLE-III calibrated photometry and 
shallow survey photometry in the {\it I}- and {\it V}-band for common stars
from the central bar subfield LMC100.1. {\it Upper panels} show the
difference as a function of $X$ image coordinate (N--S direction) while the
{\it lower panels} as a function of $Y$ coordinate (E--W direction).}
\end{figure}
\begin{figure}[p]
\includegraphics[width=9.5cm]{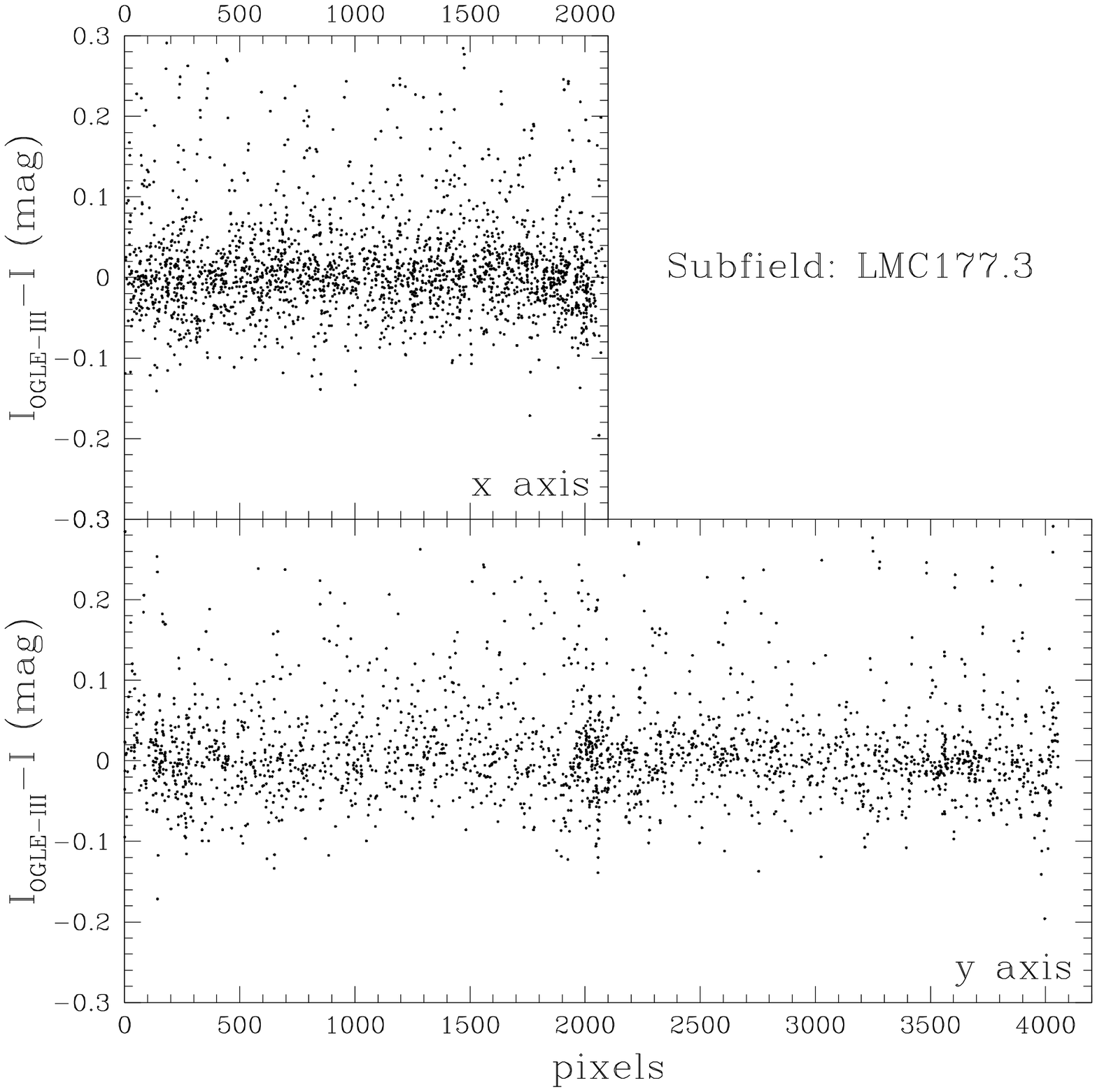}
\includegraphics[width=9.5cm]{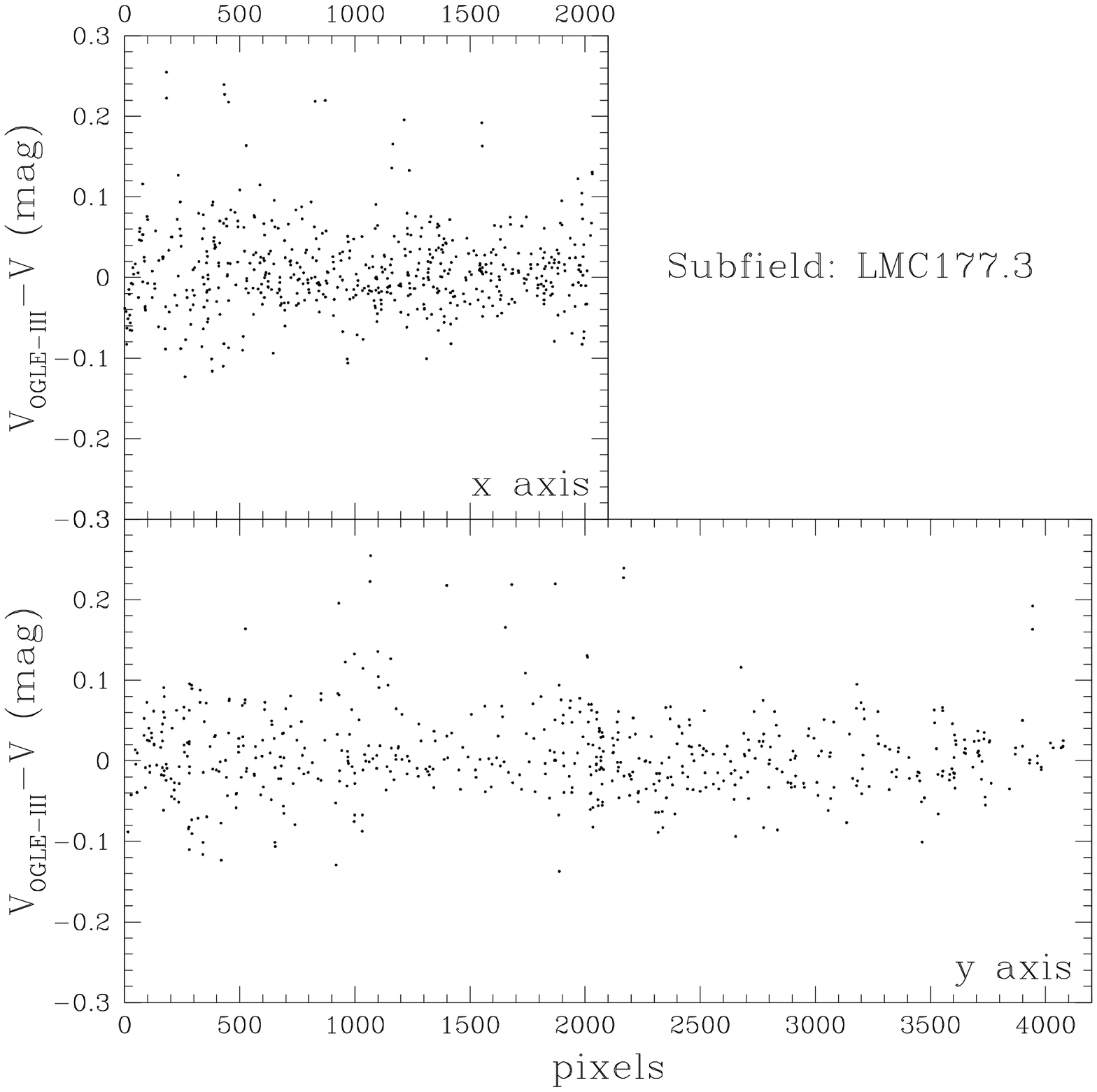}
\FigCap{Same as Fig.~4 for the LMC177.3 subfield.}
\end{figure}

The transformation of pixel coordinates to equatorial system was performed
using equations derived through procedure described in Udalski \etal
(2008a).

\Section{Construction of Photometric Maps}
The photometric databases were constructed separately for {\it I}- and {\it
V}-band data thus the first step was cross-identification between objects
registered in {\it I}- and {\it V}-band. A simple radius search was used
and closest counterpart was selected if separation was less than 2.5
pixels, what corresponds to 0\zdot\arcs65. Upper separation limit was
selected in arbitrary way after calculating minimal distances between
neighboring stars on single images.

For objects having at least 5 observations in each band 5$\sigma$ deviating
points were removed and magnitudes of all the remaining observational
points were averaged. Next, the color term correction was applied for each
object average magnitude according to the transformation equations and
color term coefficients presented in Udalski \etal (2008a). During the
conversion to the standard system the following formulae
\begin{eqnarray*}
(V-I)&=&\frac{(\vv-i)}{(1-e_V+e_I)},\\
    I&=&i+e_I\cdot(V-I),\\
    V&=&\vv+e_V\cdot(V-I)
\end{eqnarray*}
were used, where $\vv$ and $i$ are instrumental magnitudes, while $e_V$ and
$e_I$ are color term coefficients for corresponding filters.

As it was discussed in earlier papers (for example Udalski \etal 2002), the
original calibration of the OGLE-III photometry was slightly inaccurate for
very red or heavily reddened objects mainly due to the fact that the {\it
I}-band filter used during the third phase of the project was not in full
conformity with the standard Kron-Cousins definition (Landolt 1992). It
showed quite wide transparency ``wing'' in the infrared part of the
spectrum. Fortunately the OGLE-IV {\it I}-band filter can be practically
considered the standard one so it was possible to compare red stars
photometry with OGLE-III data and derive appropriate correction formula
(Szymañski \etal 2011):
$$\Delta I~= -0.033918+0.016361(V-I)+0.004167(V-I)^2.$$
We used the above relation for correction of {\it I}-band photometry and
$(V-I)$ color index for all objects with $(V-I)>1.5$ mag.

In the catalog we included interstellar extinction values obtained by
Pejcha and Stanek (2009) based on RR Lyr variables and Haschke \etal
(2012) based on red clump stars.

Table~2 presents the first 25 lines of the LMC photometric maps.  The
columns contain following data: (1)~subfield designation, (2)~ID~number,
(3,4)~equatorial coordinates J2000.0, (5,6)~corresponding $X,Y$ pixel
coordinates on the standard OGLE-III {\it I}-band reference image,
(7,8,9)~photometry: {\it I}, {\it V}, $V-I$, (10,11)~number of points for
average magnitude in {\it I}- and {\it V}-bands, (12,13)~dispersion of
magnitudes\break
\begin{landscape}
\renewcommand{\TableFont}{\scriptsize}
\renewcommand{\arraystretch}{1.05} 
\MakeTableSep{
c@{\hspace{6pt}}
r@{\hspace{7pt}}
c@{\hspace{3pt}}
c@{\hspace{5pt}}
r@{\hspace{5pt}}
r@{\hspace{5pt}}
c@{\hspace{5pt}}
c@{\hspace{5pt}}
c@{\hspace{5pt}}
c@{\hspace{5pt}}
c@{\hspace{5pt}}
c@{\hspace{5pt}}
c@{\hspace{1pt}}
c@{\hspace{-2pt}}
c@{\hspace{2pt}}|
c|
c@{\hspace{2pt}}
c|
l}
{24cm}{The first 25 lines of the LMC photometric maps based on the OGLE shallow survey}
{\hline
\noalign{\vskip3pt}
Field & ID & RA & DEC
& \multicolumn{1}{c}{$x$} & \multicolumn{1}{c}{$y$} & {\it I} & {\it V} & $V-I$ & 
$N_I$ & $N_V$ & $\sigma_I$ & $\sigma_V$ &
$N_I^{\rm bad}$&  &
$E_{\rm RRLYR}(V{-}I)$ & $E_{\rm RC}(V{-}I)$ & 
$A_I$ & ${\rm ID}_{\rm OGLEIII}$\\
& & (J2000.0) & (J2000.0) & & & [mag] & [mag] & [mag] & &&  [mag] &  
[mag]&&$N_V^{\rm bad}$&[mag]&  [mag]& [mag] &\\
\noalign{\vskip3pt}
\hline
\noalign{\vskip3pt}
LMC100.1 & s1  & 5\uph19\upm53.3\ups & $-69\arcd27\arcm34\arcs$ & 1373.3 & 1016.2 & 10.383 & 12.346 &  1.964 &  64 & 59 & 0.095 & 0.183 & 0 & 0 & 0.125 & 0.020 & 0.028 & \hspace{1cm}--\\
LMC100.1 & s2  & 5\uph20\upm16.2\ups & $-69\arcd30\arcm04\arcs$ &  792.0 & 1473.3 & 10.510 & 11.430 &  0.921 &  64 & 56 & 0.010 & 0.017 & 0 & 0 & 0.101 & 0.030 & 0.042 & \hspace{1cm}--\\
LMC100.1 & s3  & 5\uph20\upm38.3\ups & $-69\arcd30\arcm05\arcs$ &  784.3 & 1919.4 & 10.960 & 11.516 &  0.555 &  65 & 61 & 0.011 & 0.014 & 0 & 0 & 0.101 & 0.030 & 0.042 & \hspace{1cm}--\\
LMC100.1 & s4  & 5\uph19\upm32.1\ups & $-69\arcd25\arcm33\arcs$ & 1837.1 &  592.2 & 11.150 & 13.181 &  2.031 &  42 & 46 & 0.033 & 0.074 & 0 & 0 & 0.125 & 0.030 & 0.042 & \hspace{1cm}--\\
LMC100.1 & s5  & 5\uph19\upm49.6\ups & $-69\arcd30\arcm04\arcs$ &  796.6 &  939.3 & 11.264 & 12.145 &  0.882 &  62 & 56 & 0.167 & 0.052 & 0 & 0 & 0.126 & 0.030 & 0.042 & \hspace{1cm}--\\
LMC100.1 & s6  & 5\uph19\upm56.8\ups & $-69\arcd26\arcm13\arcs$ & 1680.3 & 1089.9 & 11.378 & 11.936 &  0.558 &  64 & 58 & 0.010 & 0.016 & 0 & 0 & 0.125 & 0.020 & 0.028 & \hspace{1cm}--\\
LMC100.1 & s7  & 5\uph19\upm50.2\ups & $-69\arcd26\arcm50\arcs$ & 1542.5 &  955.9 & 11.500 & 13.410 &  1.910 &  63 & 58 & 0.019 & 0.040 & 0 & 0 & 0.125 & 0.020 & 0.028 & \hspace{1cm}--\\
LMC100.1 & s8  & 5\uph19\upm51.9\ups & $-69\arcd29\arcm15\arcs$ &  985.2 &  985.7 & 11.548 & 12.259 &  0.712 &  62 & 58 & 0.008 & 0.016 & 0 & 0 & 0.126 & 0.030 & 0.042 & \hspace{1cm}--\\
LMC100.1 & s9  & 5\uph20\upm33.8\ups & $-69\arcd32\arcm35\arcs$ &  206.9 & 1823.2 & 11.640 & 11.614 & -0.026 &  39 & 24 & 0.025 & 0.020 & 0 & 0 & 0.101 & 0.030 & 0.042 & \hspace{1cm}--\\
LMC100.1 & s10 & 5\uph19\upm56.9\ups & $-69\arcd26\arcm39\arcs$ & 1581.0 & 1091.4 & 11.760 & 12.303 &  0.542 &  64 & 58 & 0.010 & 0.016 & 0 & 0 & 0.125 & 0.020 & 0.028 & \hspace{1cm}--\\
LMC100.1 & s11 & 5\uph19\upm55.3\ups & $-69\arcd28\arcm37\arcs$ & 1131.0 & 1055.9 & 11.803 & 12.540 &  0.737 &  64 & 58 & 0.009 & 0.016 & 0 & 0 & 0.125 & 0.020 & 0.028 & \hspace{1cm}--\\
LMC100.1 & s12 & 5\uph20\upm06.2\ups & $-69\arcd30\arcm02\arcs$ &  801.7 & 1273.5 & 12.012 & 13.281 &  1.269 &  63 & 58 & 0.009 & 0.017 & 0 & 0 & 0.126 & 0.040 & 0.056 & \hspace{1cm}--\\
LMC100.1 & s13 & 5\uph19\upm24.2\ups & $-69\arcd25\arcm46\arcs$ & 1789.6 &  431.6 & 12.044 & 13.711 &  1.667 &  42 & 40 & 0.017 & 0.021 & 0 & 0 & 0.125 & 0.030 & 0.042 & \hspace{1cm}--\\
LMC100.1 & s14 & 5\uph19\upm44.4\ups & $-69\arcd25\arcm43\arcs$ & 1799.4 &  840.0 & 12.206 & 13.824 &  1.618 &  59 & 59 & 0.012 & 0.018 & 0 & 0 & 0.125 & 0.020 & 0.028 & \hspace{1cm}--\\
LMC100.1 & s15 & 5\uph19\upm18.9\ups & $-69\arcd31\arcm08\arcs$ &  554.7 &  318.7 & 12.316 & 14.120 &  1.804 &  47 & 32 & 0.018 & 0.030 & 0 & 0 & 0.126 & 0.060 & 0.085 & \hspace{1cm}--\\
LMC100.1 & s16 & 5\uph19\upm41.0\ups & $-69\arcd24\arcm45\arcs$ & 2022.7 &  772.4 & 12.388 & 14.180 &  1.792 &  46 & 51 & 0.016 & 0.028 & 0 & 0 & 0.125 & 0.020 & 0.028 & \hspace{1cm}--\\
LMC100.1 & s17 & 5\uph20\upm39.0\ups & $-69\arcd32\arcm48\arcs$ &  157.9 & 1927.7 & 12.510 & 14.211 &  1.701 &  59 & 54 & 0.018 & 0.022 & 0 & 1 & 0.101 & 0.030 & 0.042 & \hspace{1cm}--\\
LMC100.1 & s18 & 5\uph19\upm44.0\ups & $-69\arcd30\arcm53\arcs$ &  608.7 &  825.0 & 12.926 & 16.563 &  3.637 &  56 & 49 & 0.168 & 0.434 & 0 & 0 & 0.126 & 0.030 & 0.042 & \hspace{1cm}--\\
LMC100.1 & s19 & 5\uph19\upm32.5\ups & $-69\arcd30\arcm35\arcs$ &  678.9 &  592.6 & 12.849 & 15.255 &  2.406 &  54 & 55 & 0.049 & 0.090 & 0 & 0 & 0.126 & 0.040 & 0.056 & LMC100.1.21\\
LMC100.1 & s20 & 5\uph19\upm23.9\ups & $-69\arcd27\arcm11\arcs$ & 1464.5 &  424.5 & 12.906 & 14.446 &  1.541 &  59 & 56 & 0.009 & 0.013 & 0 & 0 & 0.125 & 0.030 & 0.042 & LMC100.1.75626\\
LMC100.1 & s21 & 5\uph19\upm09.1\ups & $-69\arcd25\arcm55\arcs$ & 1755.8 &  125.2 & 13.082 & 15.393 &  2.311 &  56 & 54 & 0.028 & 0.077 & 0 & 0 & 0.125 & 0.070 & 0.099 & LMC100.1.75615\\
LMC100.1 & s22 & 5\uph20\upm08.8\ups & $-69\arcd32\arcm16\arcs$ &  287.0 & 1320.4 & 13.047 & 16.914 &  3.866 &  47 & 29 & 0.140 & 0.263 & 0 & 0 & 0.126 & 0.040 & 0.056 & LMC100.1.19270\\
LMC100.1 & s23 & 5\uph19\upm27.8\ups & $-69\arcd30\arcm30\arcs$ &  697.5 &  498.3 & 12.957 & 13.975 &  1.018 &  55 & 54 & 0.098 & 0.157 & 0 & 0 & 0.126 & 0.040 & 0.056 & LMC100.1.19\\
LMC100.1 & s24 & 5\uph19\upm39.4\ups & $-69\arcd25\arcm27\arcs$ & 1859.6 &  739.1 & 12.985 & 16.778 &  3.793 &  45 & 39 & 0.290 & 0.604 & 0 & 0 & 0.125 & 0.020 & 0.028 & \hspace{1cm}--\\
LMC100.1 & s25 & 5\uph19\upm33.5\ups & $-69\arcd30\arcm18\arcs$ &  744.0 &  612.7 & 13.140 & 14.694 &  1.554 &  56 & 56 & 0.125 & 0.176 & 0 & 0 & 0.126 & 0.040 & 0.056 & LMC100.1.23\\
\hline
}
\end{landscape}
\noindent for {\it I}- and {\it V}-bands, (14,15)~number of 5$\sigma$
removed points in {\it I}- and {\it V}-bands, (16-18)~extinction values for
given object coordinates, (19)~ID number of cross-identified star in the
standard OGLE-III database.  ``+'' mark before designation in column (19)
indicates multiple cross-identification (closest star was used in Table~2).

\Section{Discussion}
The data from the shallow survey presented in this paper extend already
published, detailed OGLE-III photometric maps for Large Magellanic Cloud
into the brighter magnitudes domain. We were able to obtain photometry for
objects as bright as 9.1~mag. Fig.~6 shows positions of all stars brighter
than 14.5~mag -- the structure of the galaxy can be easily recognized.
\begin{figure}[htb]
\vglue-4mm
\hglue-7mm{\includegraphics[width=14cm]{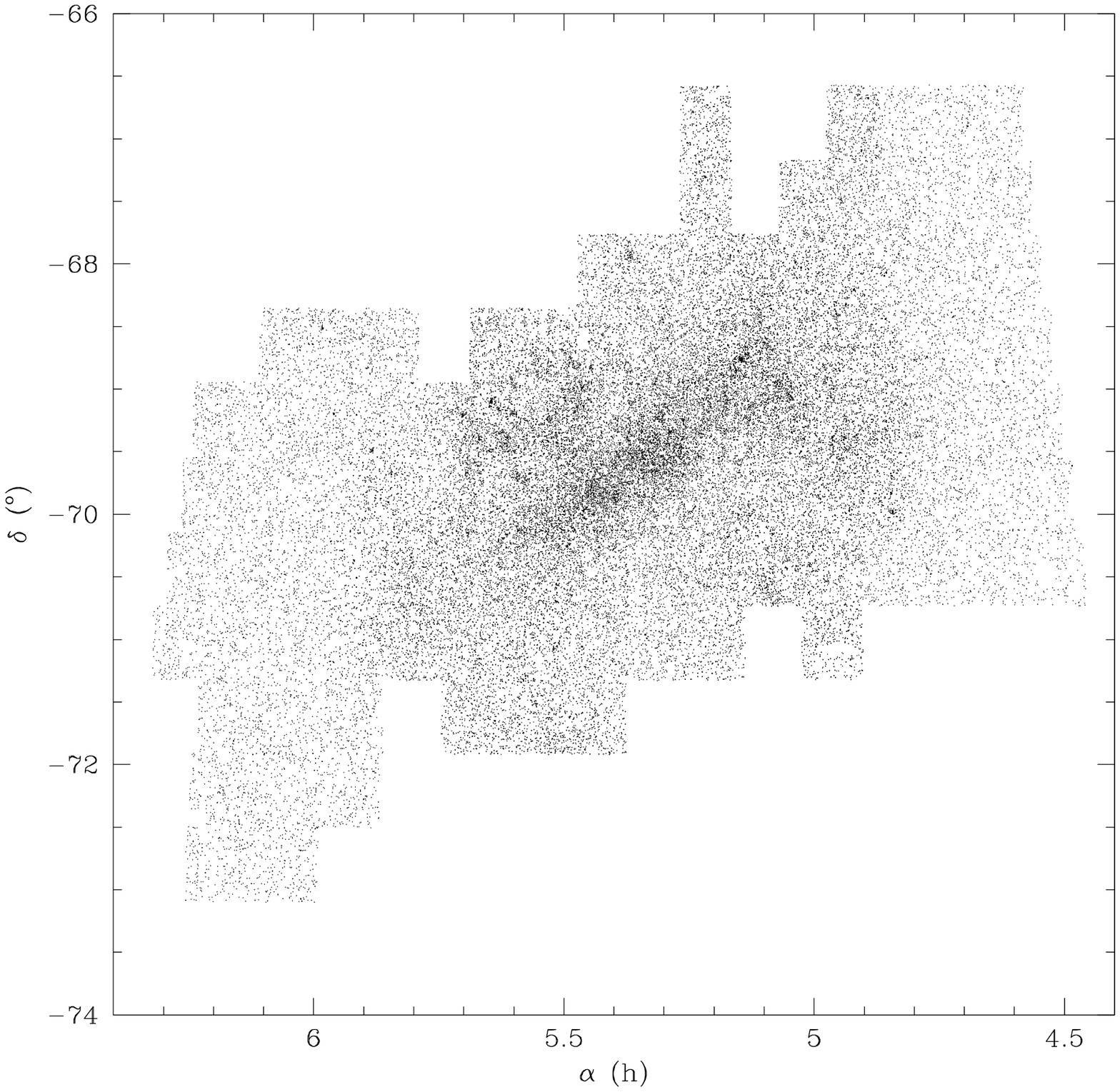}}
\vspace*{-9mm}
\FigCap{Positions of all stars brighter than 14.5~mag in the observed 
fields after removal of doubled objects from neighboring subchips/fields.}
\end{figure}
\begin{figure}[p]
\centerline{\includegraphics[width=9cm]{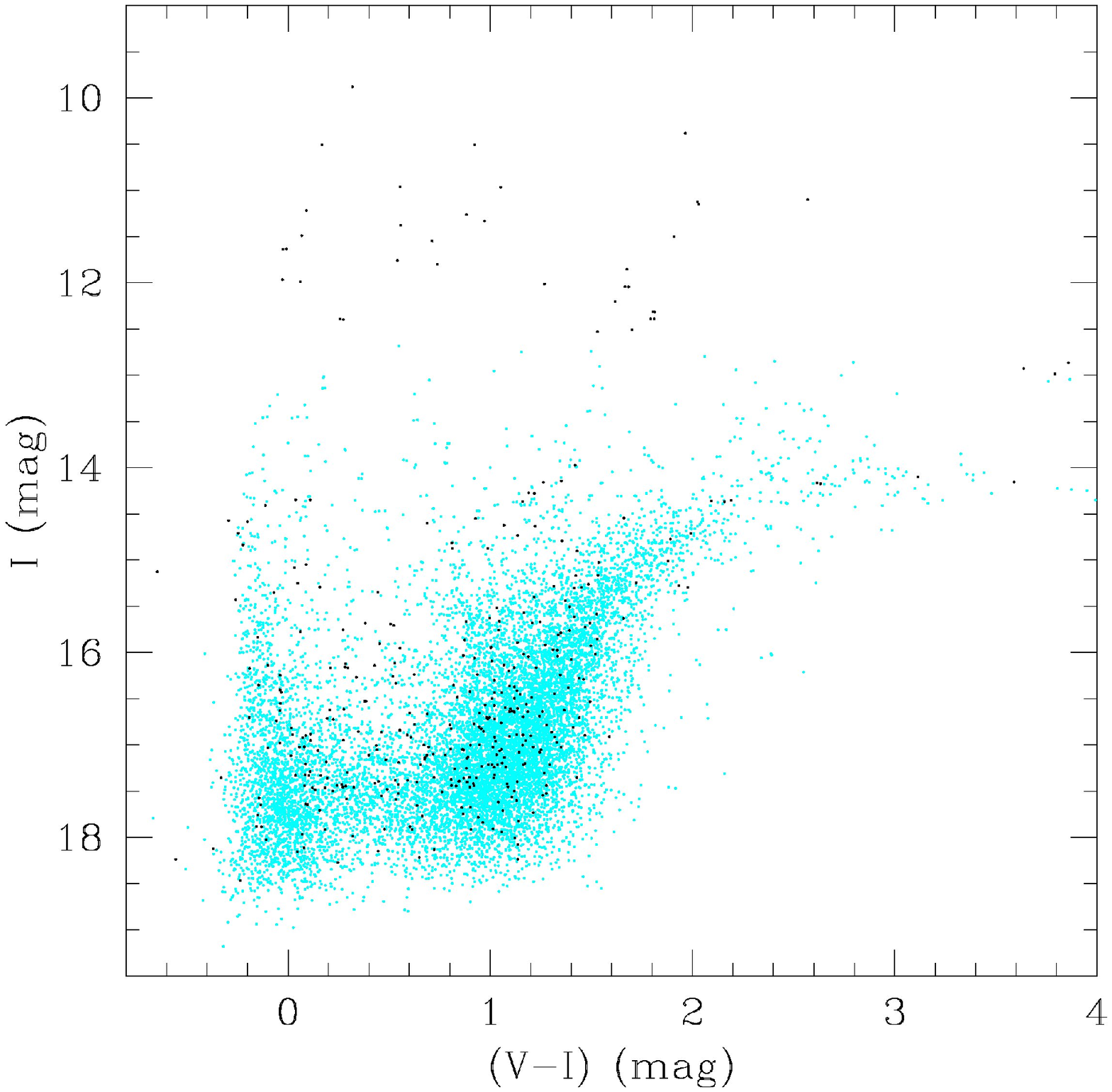}}
\FigCap{Color--magnitude diagram for the LMC100.1 subfield. The black 
points represent stars for which no corresponding objects were found in the
standard OGLE-III data.}
\vskip3mm
\centerline{\includegraphics[width=9cm]{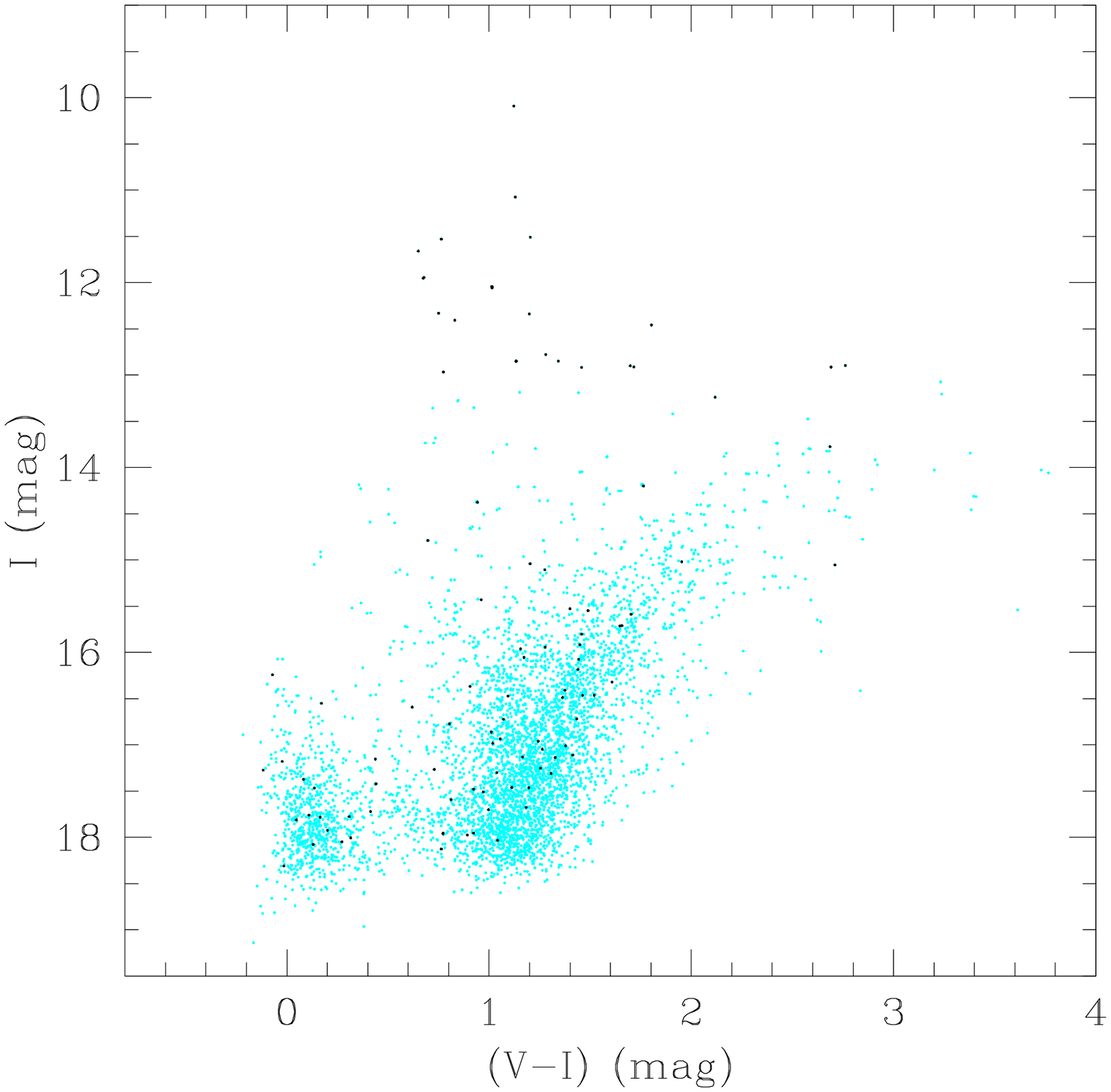}}
\FigCap{Color--magnitude diagram for the LMC185.2 subfield. The black
points represent stars for which no corresponding objects were found in the
standard OGLE-III data.}
\end{figure}

Figs.~7 and 8 present the color--magnitude diagrams based on photometric
maps for the LMC100.1 and LMC185.2 subfields. Corresponding histograms of
magnitudes in both bands are shown in Figs.~9 and~10. It is worth noting
that apart of bright stars we could also obtain completely new photometry
for many fainter stars (78\,154 objects) located close to saturated objects
in the main OGLE-III survey data. Area around such objects had to be masked
on the reference images in order to ensure high effectiveness of automatic
DIA pipelines and avoid artifact detections in this massive photometry
survey. We prepared the list with coordinates of such objects that will
allow to obtain more precise PSF photometry from regular OGLE-III images.
There is also a significant number of newly found stars (17\,480) which are
located outside OGLE-III template images. This is due to the fact that the
OGLE-III photometry is based on reference images which are stacked from
several images registered during the best photometric nights while in the
shallow survey no reference images were constructed and the complete sample
of objects detected on all frames was used in the analysis.  As it was
mentioned before, subsequent images of the same field were sometimes
considerably shifted what allowed to broaden the ``field of view'' in the
observed fields.

\renewcommand{\arraystretch}{0.97}
\renewcommand{\TableFont}{\scriptsize}
\MakeTableSep{|c|c|c@{\hspace{4pt}}r@{\hspace{4pt}}r@{\hspace{4pt}}r@{\hspace{4pt}}r||
c|c|c@{\hspace{4pt}}r@{\hspace{4pt}}r@{\hspace{4pt}}r@{\hspace{4pt}}r|}
{12.5cm}{Summary of the number of objects in the LMC fields}
{
\hline
\douprule Field & $N_{\rm OIII}$ & $N$ & $N_{\rm NEW}$ & $N_{I}$ & $N_{V}$ & $N_{VI}$
& Field & $N_{\rm OIII}$ & $N$ & $N_{\rm NEW}$ & $N_{I}$ & $N_{V}$ & $N_{VI}$\\
\hline
LMC100 & 1075711 & 78189 & 2087 & 109 & 59 & 49	 & LMC155 & 71173   & 1513  & 100   & 45  & 34  & 34 \\
LMC101 & 548707  & 33906 & 1067 & 45  & 27 & 27  & LMC156 & 72805   & 1497  & 92    & 57  & 43  & 43 \\
LMC102 & 266963  & 17906 & 694  & 40  & 48 & 38  & LMC157 & 44165   & 1652  & 165   & 58  & 43  & 43 \\
LMC103 & 833034  & 70973 & 4999 & 86  & 47 & 42  & LMC158 & 34264   & 1106  & 51    & 37  & 29  & 29 \\
LMC104 & 549136  & 34840 & 5067 & 57  & 40 & 40  & LMC159 & 234254  & 16484 & 381   & 104 & 48  & 48 \\
LMC105 & 353469  & 24324 & 419  & 106 & 39 & 39  & LMC160 & 437897  & 27916 & 466   & 172 & 72  & 72 \\
LMC107 & 226287  & 13997 & 591  & 135 & 45 & 45  & LMC161 & 787467  & 58273 & 1388  & 151 & 75  & 73 \\
LMC108 & 257801  & 14467 & 664  & 234 & 67 & 67  & LMC162 & 1171172 & 81387 & 2551  & 128 & 51  & 51 \\
LMC109 & 294145  & 18112 & 408  & 31  & 21 & 21  & LMC163 & 806912  & 48609 & 12864 & 95  & 50  & 50 \\
LMC110 & 529605  & 39037 & 2068 & 49  & 47 & 41  & LMC164 & 381352  & 25569 & 747   & 75  & 50  & 50 \\
LMC111 & 696121  & 48742 & 1997 & 77  & 32 & 32  & LMC165 & 329028  & 20376 & 509   & 80  & 40  & 40 \\
LMC112 & 752907  & 29773 & 7666 & 115 & 50 & 48	 & LMC167 & 394585  & 26472 & 628   & 135 & 49  & 49 \\
LMC113 & 561679  & 25119 & 380  & 117 & 67 & 67  & LMC168 & 636559  & 44717 & 1108  & 230 & 119 & 103 \\
LMC114 & 221912  & 15669 & 1198 & 66  & 46 & 46  & LMC169 & 1092986 & 65398 & 1273  & 155 & 63  & 63 \\
LMC117 & 527346  & 21419 & 2394 & 143 & 68 & 68	 & LMC170 & 878425  & 55419 & 1084  & 155 & 79  & 79 \\
LMC118 & 694697  & 39844 & 1824 & 48  & 50 & 39  & LMC171 & 513519  & 28705 & 2077  & 116 & 34  & 34 \\
LMC119 & 817851  & 46299 & 2083 & 186 & 60 & 60	 & LMC172 & 423099  & 22182 & 771   & 91  & 62  & 62 \\
LMC120 & 617701  & 30819 & 371  & 104 & 38 & 38  & LMC174 & 340208  & 14762 & 1638  & 130 & 52  & 52 \\
LMC121 & 442352  & 31664 & 778  & 113 & 76 & 65  & LMC175 & 497977  & 23122 & 1617  & 402 & 205 & 204 \\
LMC124 & 290411  & 18125 & 621  & 74  & 50 & 50  & LMC176 & 576984  & 22129 & 1968  & 150 & 85  & 84 \\
LMC125 & 357288  & 21252 & 640  & 70  & 50 & 33  & LMC177 & 817966  & 26121 & 613   & 108 & 63  & 63 \\
LMC126 & 530735  & 30877 & 897  & 31  & 59 & 25  & LMC178 & 471477  & 18525 & 2964  & 91  & 52  & 52 \\
LMC127 & 547901  & 35676 & 2473 & 166 & 57 & 57	 & LMC179 & 296759  & 13925 & 550   & 70  & 51  & 51 \\
LMC128 & 406243  & 25285 & 675  & 30  & 51 & 26  & LMC183 & 361838  & 18671 & 1566  & 156 & 71  & 71 \\
LMC129 & 304616  & 20516 & 1387 & 68  & 46 & 36  & LMC184 & 486666  & 24001 & 524   & 129 & 65  & 64 \\
LMC130 & 231039  & 13949 & 1615 & 137 & 41 & 41	 & LMC185 & 630366  & 32496 & 705   & 147 & 80  & 80 \\
LMC131 & 248421  & 8279  & 364  & 136 & 42 & 42  & LMC186 & 376966  & 28750 & 2662  & 65  & 48  & 47 \\
LMC132 & 217827  & 7998  & 147  & 59  & 21 & 21  & LMC189 & 153656  & 3784  & 256   & 70  & 46  & 46 \\
LMC133 & 346558  & 16403 & 1976 & 237 & 112& 112 & LMC190 & 200564  & 5106  & 904   & 78  & 34  & 34 \\
LMC134 & 304271  & 18835 & 300  & 52  & 43 & 38  & LMC191 & 243558  & 10646 & 951   & 148 & 76  & 72 \\
LMC135 & 284846  & 20805 & 2808 & 146 & 54 & 50	 & LMC192 & 228516  & 10391 & 611   & 110 & 101 & 93 \\
LMC136 & 244490  & 19405 & 2673 & 74  & 20 & 15  & LMC193 & 138223  & 9098  & 480   & 74  & 77  & 74 \\
LMC137 & 200058  & 5965  & 539  & 61  & 24 & 24  & LMC194 & 77816   & 2941  & 69    & 30  & 34  & 27 \\
LMC138 & 117664  & 3076  & 369  & 22  & 17 & 17  & LMC195 & 43002   & 2088  & 327   & 36  & 30  & 30 \\
LMC139 & 128858  & 4141  & 226  & 36  & 24 & 24  & LMC196 & 116258  & 4373  & 108   & 18  & 19  & 16 \\
LMC140 & 188164  & 5050  & 103  & 46  & 34 & 34  & LMC197 & 92295   & 3769  & 116   & 36  & 28  & 28 \\
LMC141 & 191197  & 5660  & 247  & 54  & 36 & 35  & LMC198 & 71992   & 3488  & 223   & 56  & 37  & 37 \\
LMC142 & 225807  & 6808  & 940  & 81  & 36 & 36  & LMC199 & 63841   & 3154  & 144   & 34  & 25  & 25 \\
LMC143 & 155672  & 5329  & 330  & 52  & 24 & 24  & LMC200 & 58765   & 3013  & 193   & 30  & 24  & 24 \\
LMC144 & 116363  & 3481  & 508  & 29  & 24 & 24  & LMC201 & 91051   & 2254  & 362   & 81  & 57  & 57 \\
LMC145 & 64628   & 1902  & 222  & 18  & 19 & 17  & LMC202 & 79469   & 2158  & 175   & 62  & 55  & 55 \\
LMC146 & 86656   & 2621  & 240  & 20  & 21 & 17  & LMC203 & 71396   & 1964  & 201   & 44  & 48  & 42 \\
LMC147 & 103604  & 3105  & 167  & 43  & 31 & 29  & LMC204 & 108173  & 3085  & 118   & 72  & 48  & 48 \\
LMC148 & 110885  & 3000  & 290  & 40  & 39 & 35  & LMC205 & 82072   & 2755  & 185   & 48  & 28  & 28 \\
LMC149 & 115117  & 3448  & 268  & 32  & 31 & 28  & LMC206 & 78179   & 2357  & 364   & 43  & 43  & 41 \\
LMC150 & 96039   & 3119  & 100  & 16  & 27 & 16  & LMC207 & 70889   & 2036  & 68    & 22  & 27  & 21 \\
LMC151 & 89935   & 2403  & 170  & 35  & 22 & 22  & LMC208 & 87619   & 2046  & 98    & 69  & 54  & 54 \\
LMC152 & 46107   & 1468  & 96   & 63  & 56 & 56  & LMC209 & 64885   & 1737  & 81    & 23  & 53  & 22 \\
LMC153 & 56016   & 1565  & 92   & 49  & 40 & 40  & LMC210 & 70282   & 1610  & 97    & 8   & 28  & 7 \\
LMC154 & 65095   & 1635  & 61   & 32  & 31 & 30  & LMC211 & 61205   & 1473  & 50    & 46  & 29  & 29 \\
LMC155 & 71173   & 1513  & 100  & 45  & 34 & 34  & LMC212 & 81878   & 2017  & 54    & 19  & 23  & 18 \\
LMC156 & 72805   & 1497  & 92   & 57  & 43 & 43  & LMC213 & 52207   & 1926  & 111   & 44  & 48  & 44 \\
LMC157 & 44165   & 1652  & 165  & 58  & 43 & 43  & LMC214 & 60410   & 1661  & 108   & 23  & 23  & 22 \\
LMC158 & 34264   & 1106  & 51   & 37  & 29 & 29  & LMC215 & 60801   & 1236  & 58    & 29  & 20  & 20 \\
\hline
}
\begin{figure}[p]
\vglue-3mm
\centerline{\includegraphics[width=9.5cm]{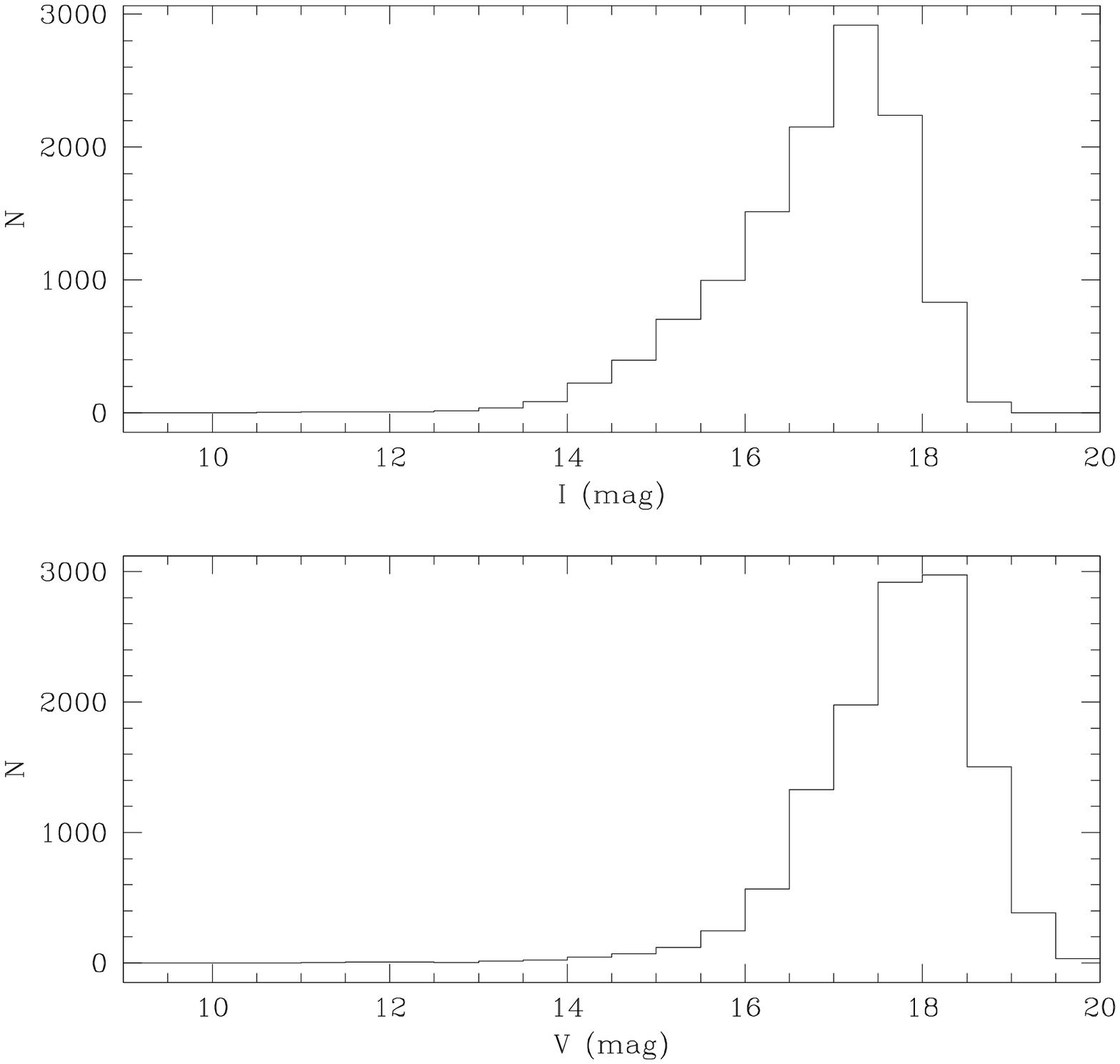}}
\vspace*{-3pt}
\FigCap{Histogram of magnitudes for the LMC100.1 subfield}
\vskip3mm
\centerline{\includegraphics[width=9.5cm]{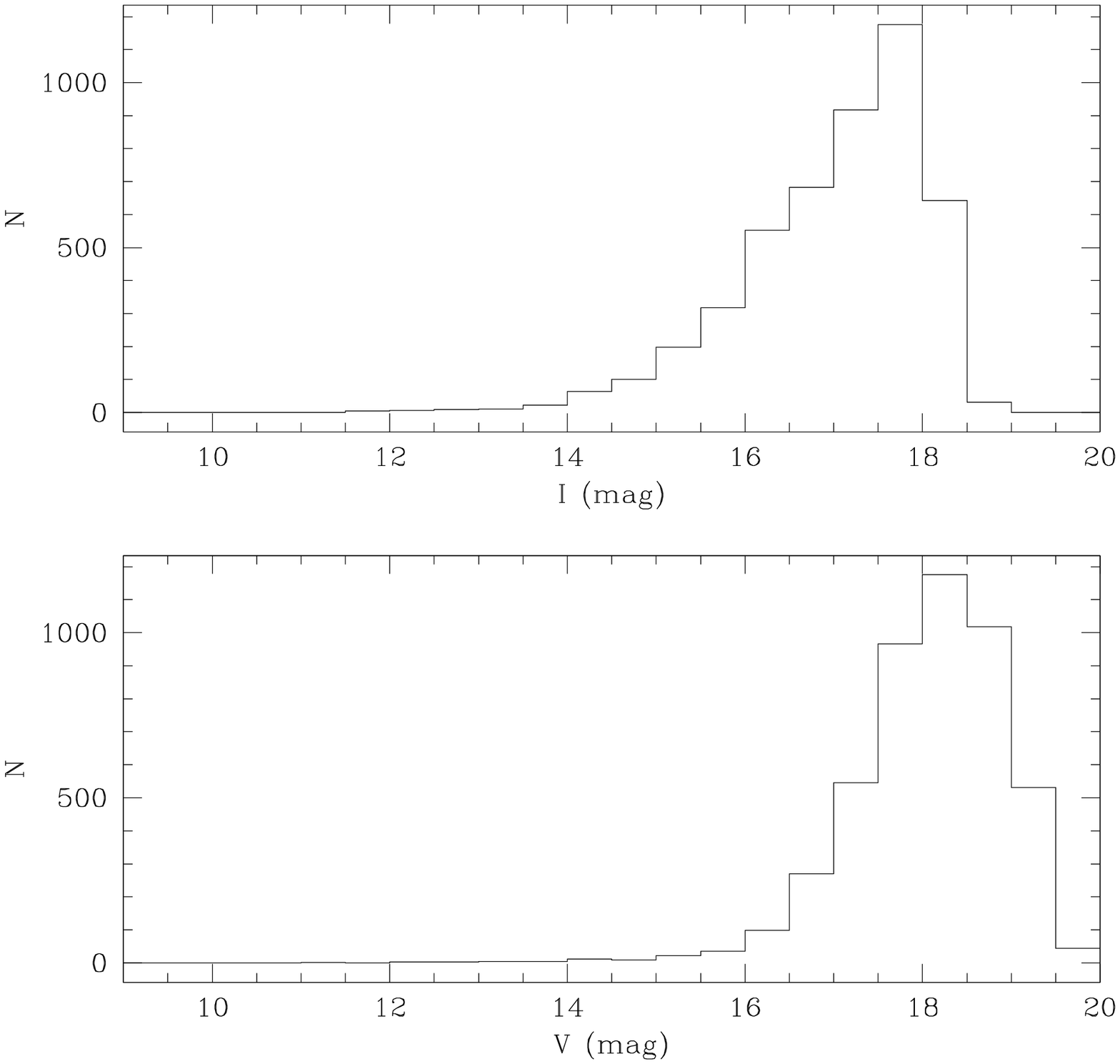}}
\vspace*{-3pt}
\FigCap{Histogram of magnitudes for the LMC185.2 subfield}
\end{figure}

Table~3 summarizes obtained photometry for all objects. In the subsequent
columns following data are presented: (1)~OGLE-III field name, (2)~number
of objects for which photometry is available in the main OGLE-III
photometric maps, (3)~total number of objects in the shallow survey
photometric maps for given field, (4)~number of new objects (not present in
the main OGLE-III photometric maps) for given field, (5)~number of new
objects brighter in the {\it I}-band than brightest star for given field in
the main OGLE-III photometric maps, (6)~same as (5) but for the {\it
V}-band, (7)~number of new objects brighter in the {\it I}-band and {\it
V}-band than brightest stars in the corresponding bands for given field in
the main OGLE-III photometric maps.

To show quality of obtained PSF photometry we present comparison of light
curves for some unblended Cepheid variables to the standard OGLE-III and
ASAS (Pojmañski 1997 and 2002) data (Figs. 11--17). As one can see despite
the weather was much worse than during usual observational conditions the
quality of photometry turned out to be good. Obviously, for some
non-isolated stars one can expect significant blending due to blurred
stellar profiles.

\begin{figure}[p]
\vglue-11mm
\hglue-1cm{\includegraphics[width=14.5cm, height=20.5cm]{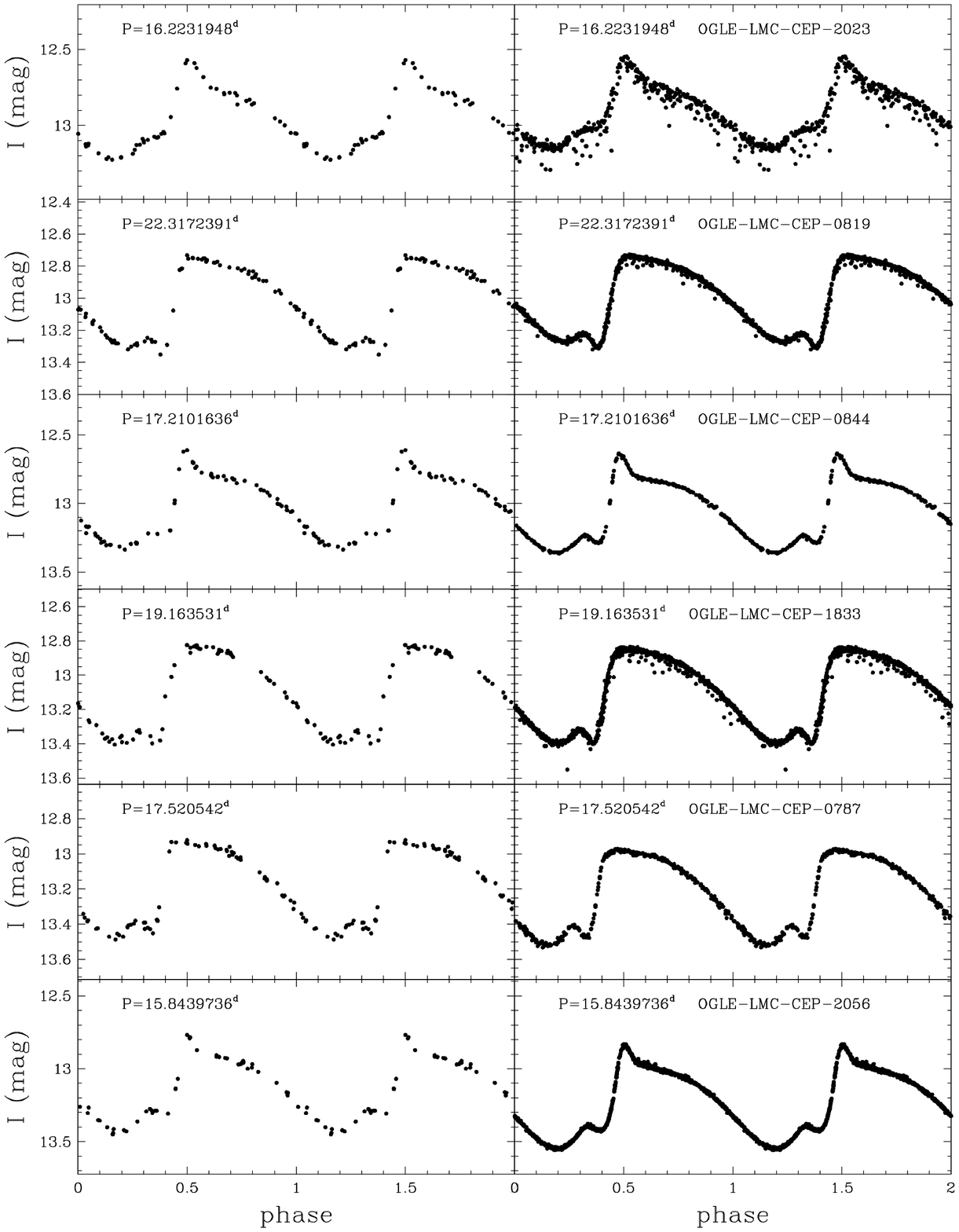}}
\vspace*{-1.3cm}
\FigCap{Comparison of shallow survey {\it I}-band light curves of 
classical Cepheids in the LMC ({\it left panels}) with the standard
OGLE-III data ({\it right panels}).}
\end{figure}
\begin{figure}[p]
\vglue-11mm
\hglue-1cm{\includegraphics[width=14.5cm, height=20.5cm]{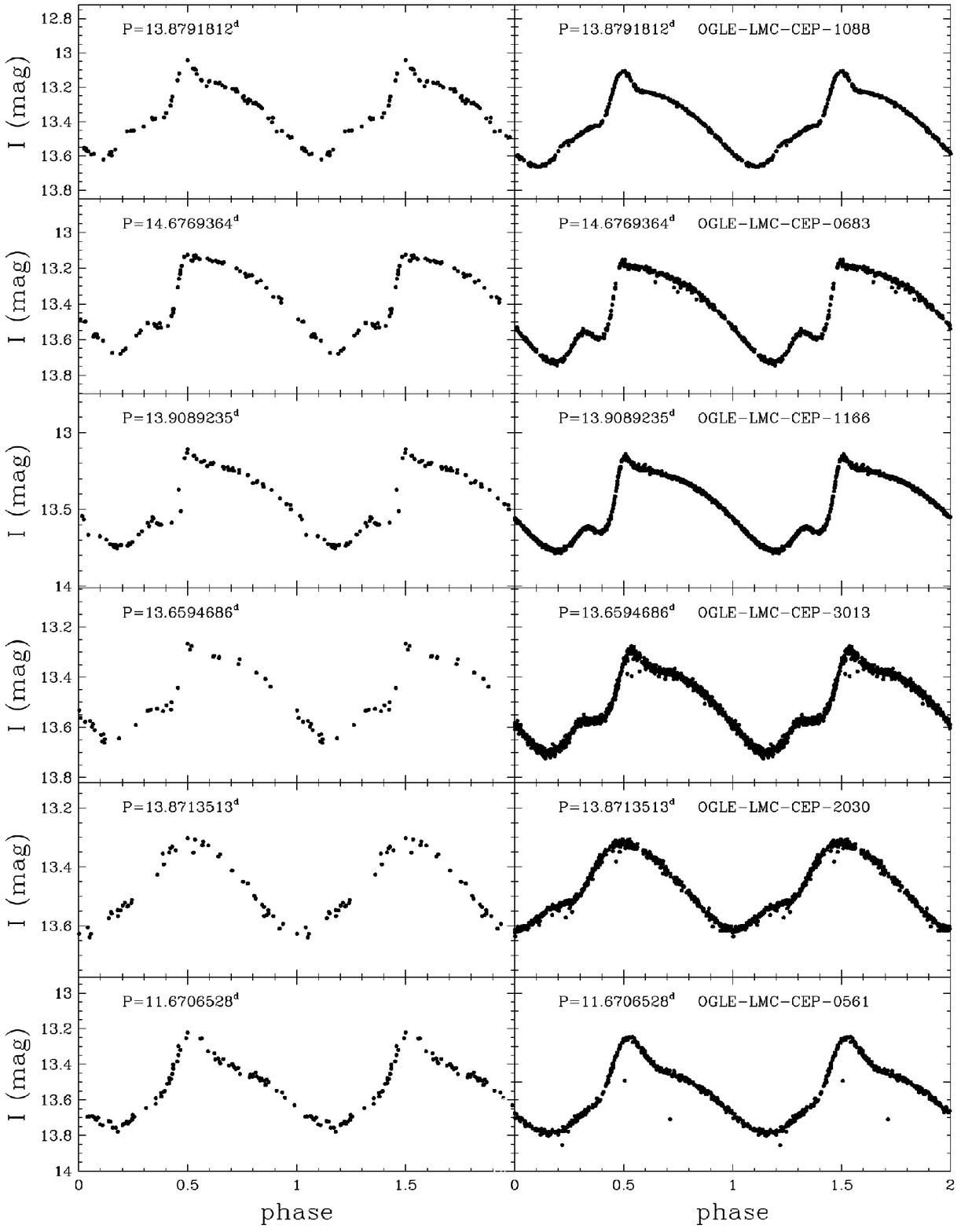}}
\vspace*{-1.3cm}
\FigCap{Same as in Fig.~11.}
\end{figure}
\begin{figure}[p]
\vglue-11mm
\hglue-1cm{\includegraphics[width=14.5cm, height=20.5cm]{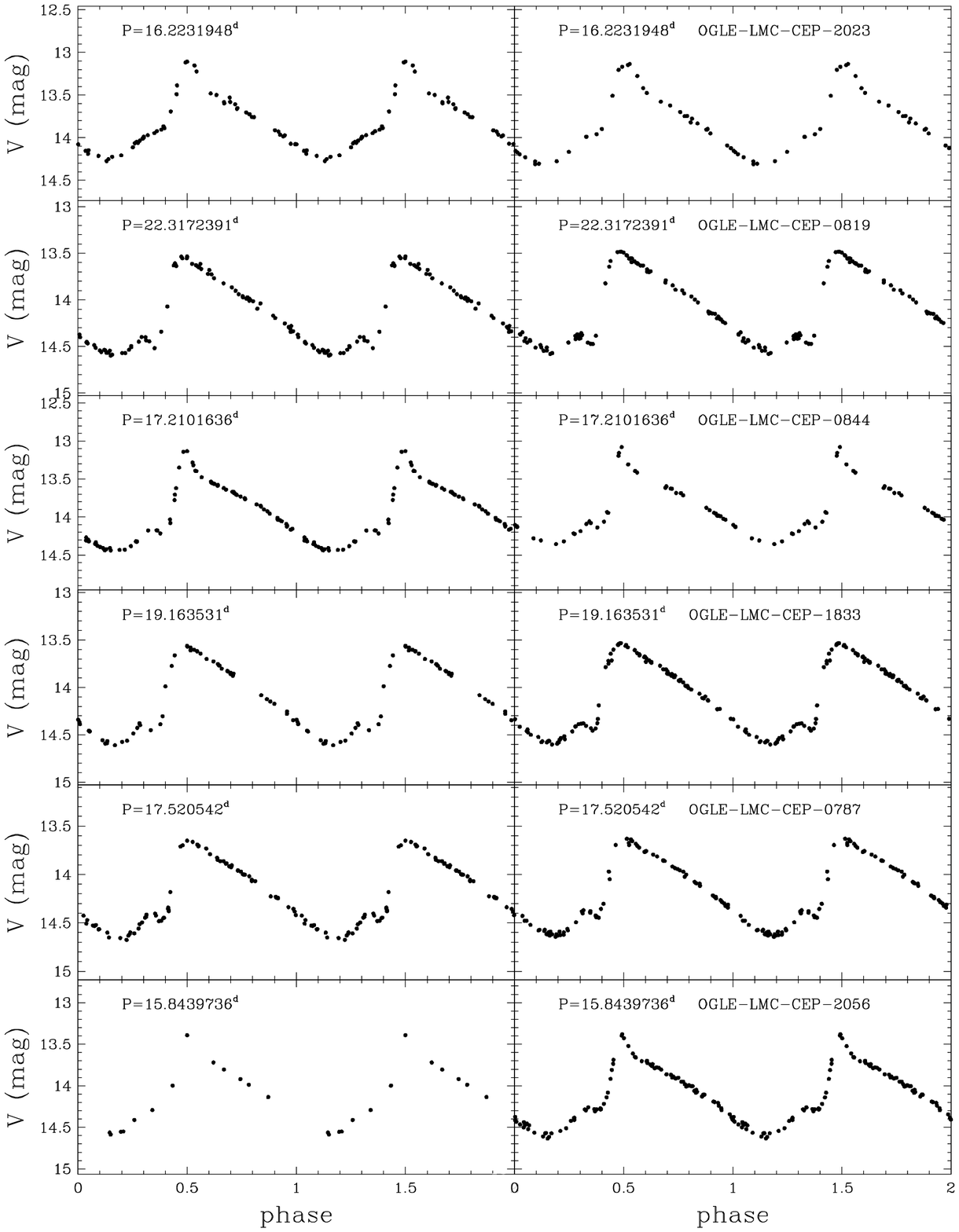}}
\vspace*{-1.3cm}
\FigCap{Comparison of shallow survey {\it V}-band light curves of 
classical Cepheids in the LMC ({\it left panels}) with the standard
OGLE-III data ({\it right panels}) for the same objects as in Fig.~11.}
\end{figure}
\begin{figure}[p]
\vglue-11mm
\hglue-1cm{\includegraphics[width=14.5cm, height=20.5cm]{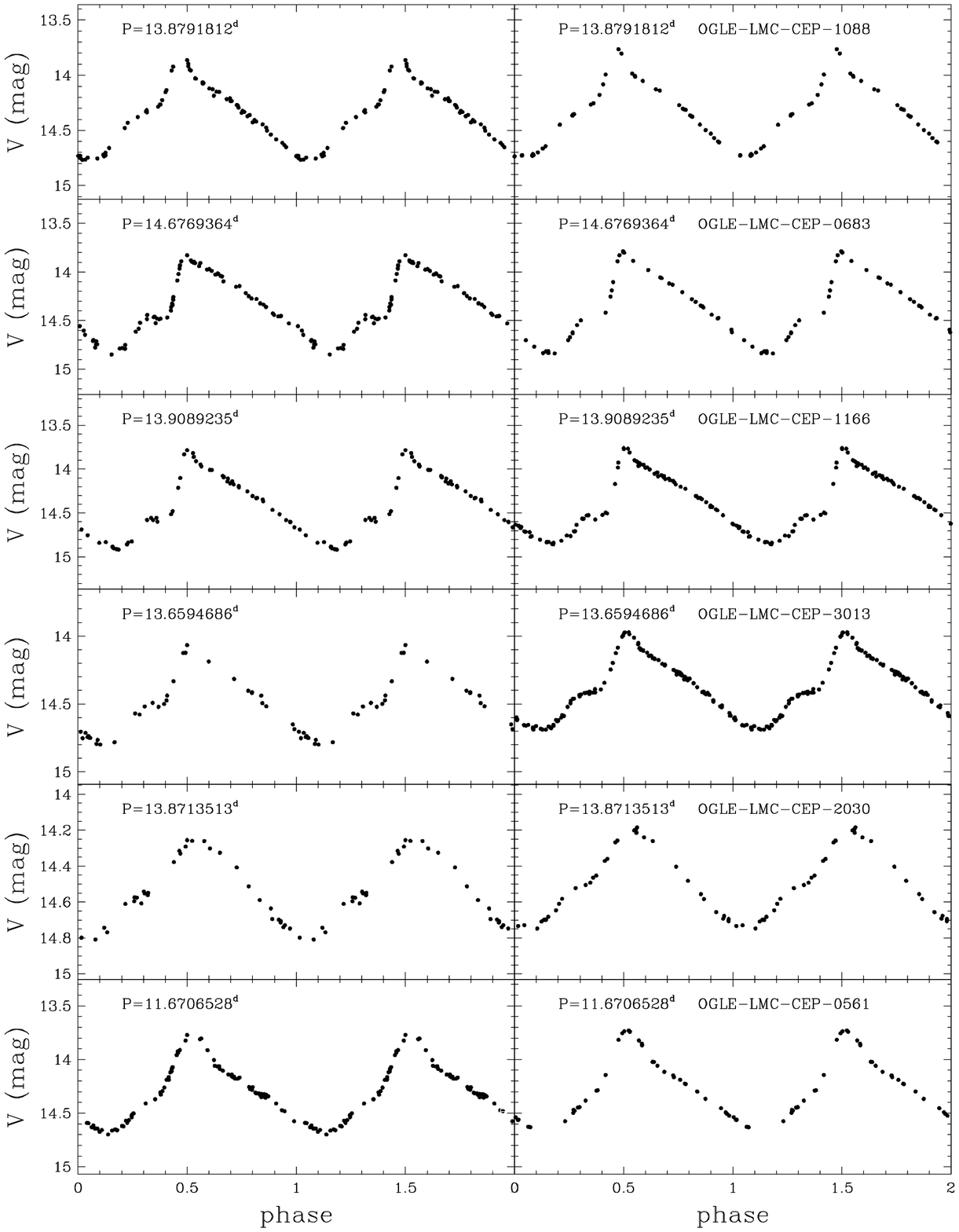}}
\vspace*{-1.3cm}
\FigCap{Comparison of shallow survey {\it V}-band light curves of 
classical Cepheids in the LMC ({\it left panels}) with the standard
OGLE-III data ({\it right panels}) for the same objects as in Fig.~12.}
\end{figure}
\begin{figure}[p]
\vglue-11mm
\hglue-1cm{\includegraphics[width=14.5cm, height=20.5cm]{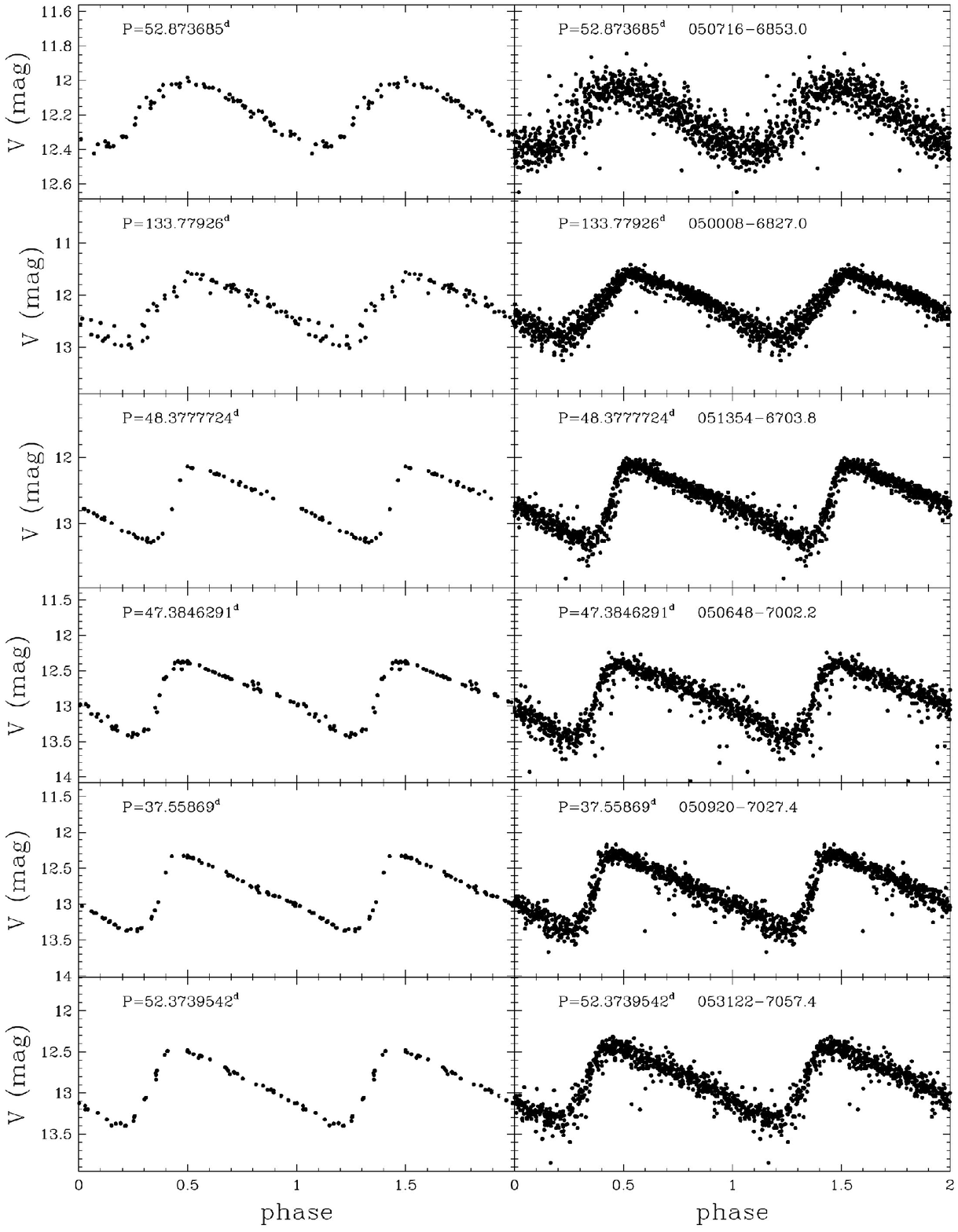}}
\vspace*{-1.3cm}
\FigCap{Comparison of light curves of classical Cepheids in the LMC 
in the {\it V}-band with the ASAS data ({\it right panels}).}
\end{figure}
\begin{figure}[p]
\vglue-11mm
\hglue-1cm{\includegraphics[width=14.5cm, height=20.5cm]{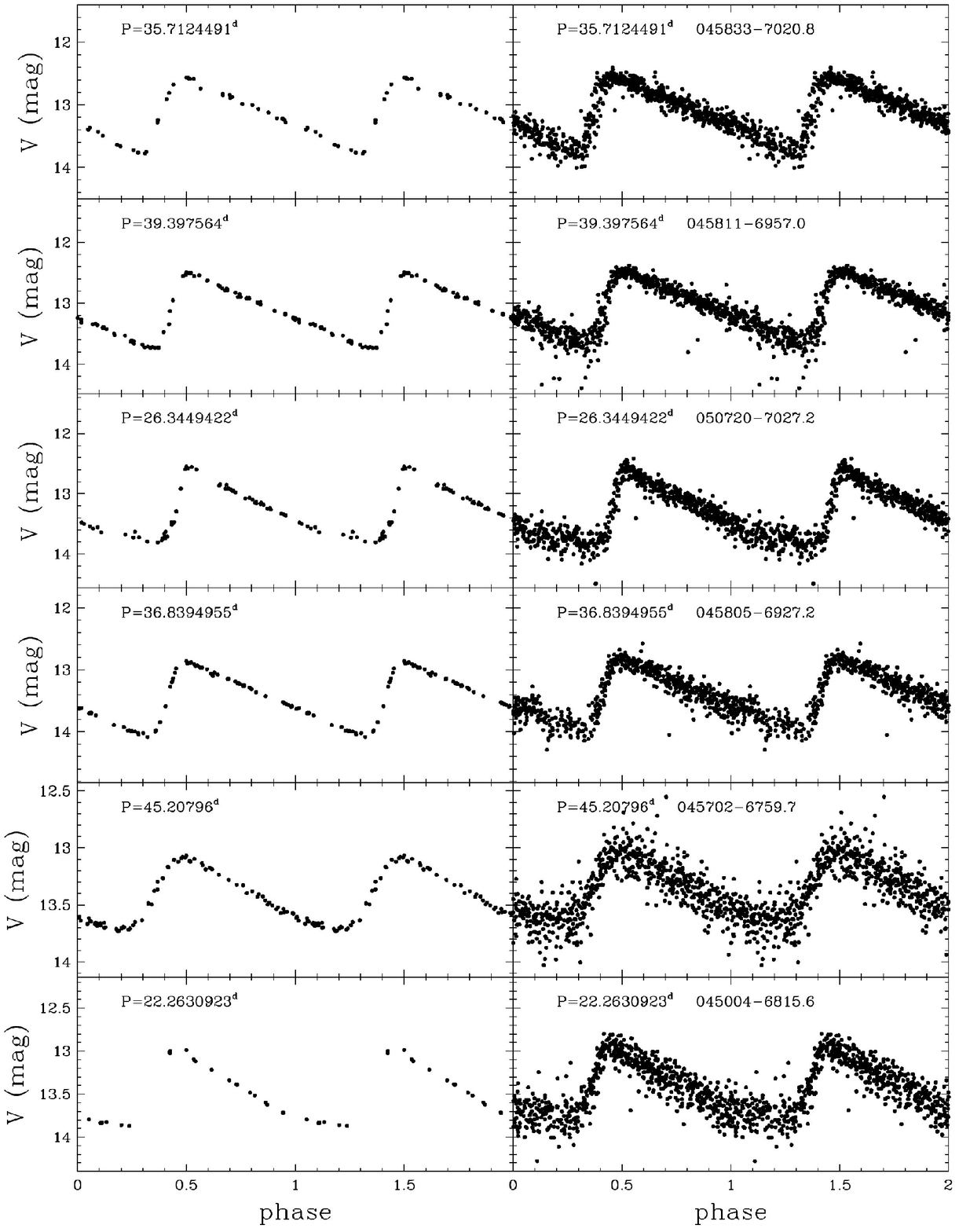}}
\vspace*{-1.3cm}
\FigCap{Same as in Fig.~15.}
\end{figure}
\begin{figure}[p]
\vglue-11mm
\hglue-1cm{\includegraphics[width=14.5cm, height=20.5cm]{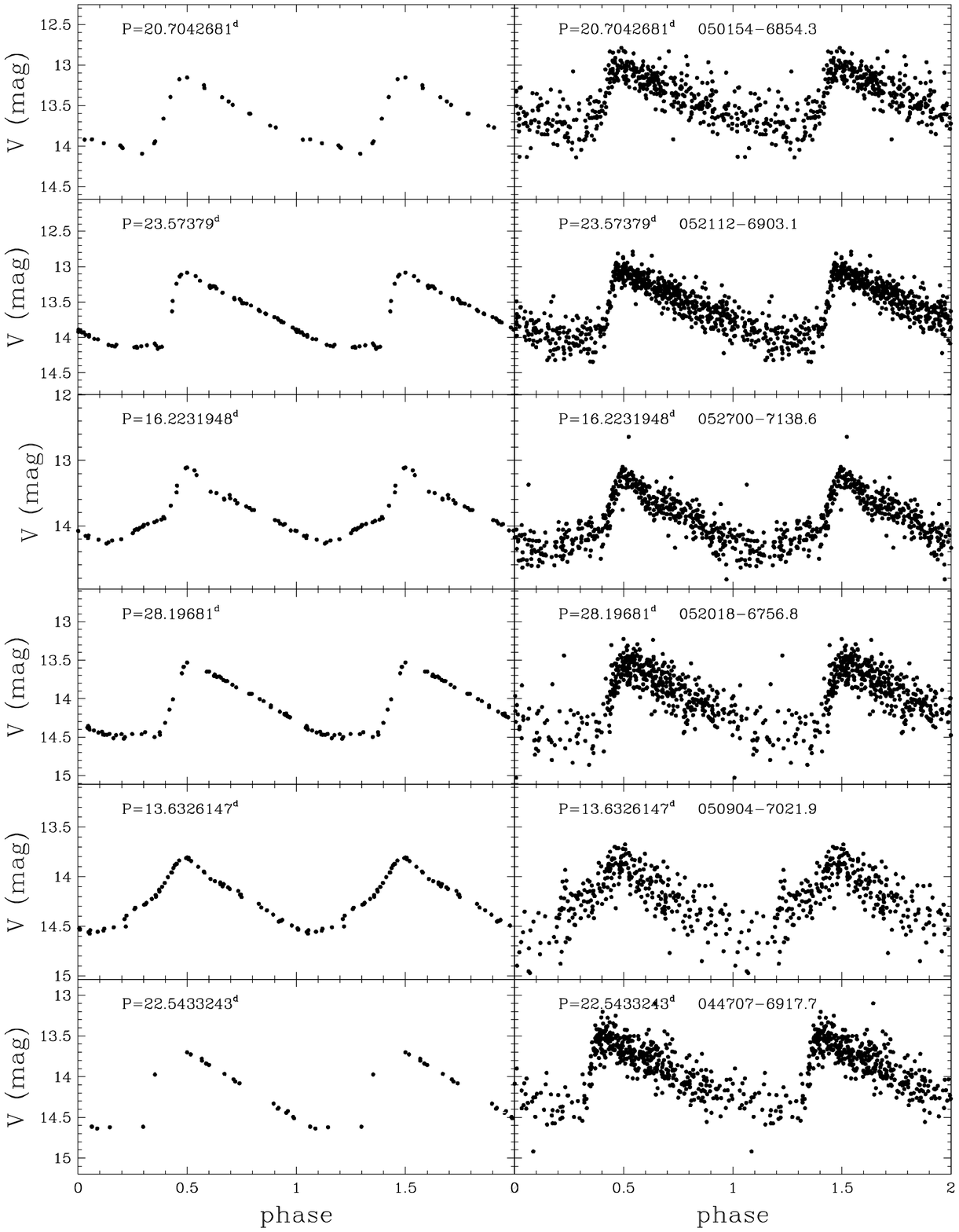}}
\vspace*{-1.3cm}
\FigCap{Same as in Fig.~15.}
\end{figure}
Beside the OGLE LMC mapping several other photometric surveys have been
conducted for the LMC. They based usually on one/few-epoch data what
limited possible apllications of these studies. Cioni \etal (2000)
presented point sources catalog (ca.\ 1.3 million objects) from the DENIS
survey carried out using {\it I}, {\it J} and $K_S$ filters and consisting
of objects for which $10.5~{\rm mag}<I<18$~mag. {\it UBVR} catalog of
179\,655 stars within $V=12\div18$~mag range was presented by Massey
(2002). However in that case the galaxy was not fully covered. Finally,
Zaritsky \etal (2004) used Johnson {\it U}, {\it B}, {\it V} and Gunn {\it
i} filters to obtain photometry for 24 million stars in the LMC with
brightness of $13.5\div20$~mag in the {\it V}-band.

\newpage
In the forthcoming paper we will present results of variability analysis
performed for all new objects. For the Small Magellanic Cloud complete
sets of short-exposure images were also collected and we plan to publish
similar data for this galaxy.

\Section{Data Availability}
The photometric maps for bright objects in the LMC are available to the
astronomical community from the OGLE Internet Archive:

\begin{center}
{\it http://ogle.astrouw.edu.pl

ftp://ftp.astrouw.edu.pl/ogle3/maps/lmc\_bright/}
\end{center}

\Acknow{The OGLE project has received funding from the European Research 
Council under the European Community's Seventh Framework Programme
(FP7/2007-2013)/ERC grant agreement no. 246678 to AU. This paper was
partially supported by Polish grant N~N203 510038. We gratefully
acknowledge financial support for this work from the Chilean Center for
Astrophysics FONDAP 15010003, and from the BASAL Centro de Astrofisica y
Tecnologias Afines (CATA) PFB-06/2007.}

\end{document}